\documentstyle[pre,aps]{revtex}

\begin{document}
\renewcommand{\theequation}{\arabic{section}.\arabic{equation}}

\draft
\title{
Shape Changes of Deformable Spherical Membranes with $n$-atic Order
}
\author{Jeong-Man Park}
\address{
Department of Physics, University of Pennsylvania, 
Philadelphia, PA 19104
}
\maketitle

\begin{abstract}
We present the existence of the Kosterlitz-Thouless (KT) transition
for $n$-atic tangent-plane order on a deformable spherical surface
and investigate the development of quasi-long range $n$-atic order 
and the continuous
shape changes of a deformable spherical surface below the KT
transition in the low temperature (mean field) limit.
The $n$-atic order parameter $\psi= \exp[in\Theta]$
describes, respectively, vector, nematic, and hexatic order for
$n=1,2,$ and 6. We derive a Coulomb gas Hamiltonian to describe it.
We then convert it into the sine-Gordon Hamiltonian and find a linear
coupling between a scalar field and the Gaussian curvature on the
fluctuating spherical surface. After integrating over the shape
fluctuations, we obtain the massive sine-Gordon Hamiltonian, where
the interaction between vortices is screened.
We find, for $n^{2}K_{n}/\kappa \gg 1/4$ where $K_{n}$ and $\kappa$
are $n$-atic and bending rigidity, respectively, the KT transition
is supressed altogether. On the other hand, if $n^{2}K_{n}/\kappa \ll 1/4$,
there is an effective KT transition.
In the ordered phase, tangent-plane $n$-atic order expels the Gaussian 
curvature. 
In addition, the total vorticity of orientational order on a
surface of genus zero is two.  Thus, the ordered phase of an $n$-atic
on such a surface
will have $2n$ vortices of strength $1/n$, $2n$ zeros in its order
parameter, and a nonspherical equilibrium shape.  
Our calculations are based on a
phenomenological model with a gauge-like coupling between $\psi$ and
curvature, and our analysis follows closely the Abrikosov treatment of a
type II superconductor just below $H_{c2}$.
\end{abstract}
\par
\section{Introduction}
When surfactant molecules, or amphiphiles which consist of molecules that
combine both polar and non-polar parts, are dissolved in a single solvent
such as water, these molecules tend to form bilayer membranes where
the hydrocarbon parts of each monolayer are aggregated in the middle of
the bilayer to reduce the contact between the water and the non-polar parts
of the molecule due to the hydrophobic interactions of the hydrocarbon chains.
These membranes are not covalenly bonded, but rather stabilized by the
weaker hydrophobic interaction. These membranes can have characteristic 
sizes which may be much larger than those of a single molecule. The sizes and
shapes of these membranes can change as a function of temperature, salinity,
and/or surfactant concentration. Depending on the physical conditions of
the system, these membranes can form random extended surfaces, regular
periodic structures, or closed vesicles separating an interior from an 
exterior \cite{Nelson-89}. 
Furthermore, some bilayer membranes are prototypes of
biological systems, although it should be noted that true biological membranes
have several components and greater complexity.
\par
In a wide class of membranes,
molecules move freely within the membrane forming a $2$-dimensional
fluid offering little resistance to changes in membrane shape. 
Such membranes are physical examples of random surfaces, which
can undergo violent shape changes.
Molecules in membranes can also exhibit
varying degrees of orientational and positional order 
including tilt(smectic-C), hexatic, and crystalline orderings 
similar to those found in free standing liquid crystal
films \cite{Luzzatti-68,Tardieu-73,Smith-88}.  
These ordered membranes provide fascinating laboratories for
the study of the coupling between order and geometry, analogous in many ways
to the coupling between matter and geometry in general relativity.  The
underlying cause of this coupling is as follows.  A field 
describing orientational order cannot be parallel everywhere if it is
forced to lie on a surface, such as
that of a sphere, that curves in two directions and thus has nonzero
Gaussian curvature.  It could lower its energy
by flattening the surface.  An additional complication arises when order
develops on closed surfaces.  A closed surface can be classified according
to its genus $g$: the number of handles.
Orientational order on a closed surface necessarily has topological
defects (vortices) with total strength (vorticity) equal to the Euler
characteristic $2(1-g)$ of the surface \cite{Spivak-79}.  
Tangent-plane order on a
sphere (torus) will have vorticity $2$ $(0)$, since a sphere (torus)
has genus $0$ ($1$), respectively. 
The continuous development of vector order on a deformable surface of genus
zero will be accompanied by a continuous change from spherical to
ellipsoidal shape \cite{Park-92}. Since vortices
are energetically costly, it may be favorable for a closed physical
membrane to transform into an open cylindrical structure when 
tangent-plane order develops in response to changes in temperature or 
other control variables \cite{MacKintosh-91}.  
Indeed, there are a number of experimental 
examples of shape changes that may
be explained by the development of tangent-plane order.
\par
Orientational orders for smectic-C and hexatic liquid crystals are described
by an $n$-atic order parameter $\psi = \langle \exp [i n \theta
] \rangle$ with $n=1$ and $n=6$, respectively.
In this paper, we present the existence of the KT transition of
$n$-atic order on a closed surface of genus zero and investigate 
the development
of $n$-atic order below the KT transition temperature and the concomitant
change in shape from spherical to non-spherical.  An $n$-atic order
parameter can have vortices of strength $1/n$, and, since it is generally
favorable to form vortices of minimum strength, we expect $2n$ maximally
separated vortices of strength $1/n$ to be present in the ordered phase (low
temperature phase).  Thus
for $n=1$, we expect two antipodal vortices, and respectively
for $n=2$, $n=3$ and $n=6$, we expect 
vortices at the vertices of a tetrahedron, an octahedron, and an icosahedron.
Indeed calculations on a
rigid sphere confirm this conjecture \cite{Lubensky-92}.
We also expect the vesicle
shape to change from spherical to ellipsoidal, tetrahedral, octahedral, or
icosahedral in the four cases above. 
\par
Our calculations are based on a phenomenological Hamiltonian for a
complex order parameter field whose coupling to shape occurs via a
covariant derivative and via changes in the metric tensor \cite{Nelson-87}.  
The model is
almost identical to the Landau-Ginzburg theory of superconductivity except
that vorticity is fixed by surface topology rather than energetically
determined by an external magnetic field.  The ordering transition we find
is very similar to the transition from a normal metal to the Abrikosov
vortex lattice in a superconductor, and indeed our analysis follows
very closely that of Abrikosov \cite{Abrikosov-57}.  
We find the order parameter, which is
necessarily spatially inhomogeneous because of the topological constraint on
the total vorticity, from among a highly degenerate
set of functions that diagonalize the harmonic Hamiltonian on a rigid
sphere.  This degenerate set has exactly $2n$ zeros at arbitrary positions
on the sphere and is very similar in form to fractional quantum Hall wave
functions \cite{Haldane-83}.
\par
This paper is organized as follows. In Sec. II, we review fluid membranes 
in some detail and describe their free energy as a function of 
curvature. We introduce a tangent-plane orientational order parameter
and its coupling to shape changes in Sec. III. The model Hamiltonian for
a deformable spherical membrane with $n$-atic order is presented in the
normal gauge representation and the existence of the KT transition
is described in Sec. IV. Minimization scheme and
the calculated shapes for temperatures below the KT transition temperature
to the ordered $n$-atic phase for $n=1,2,3,4$, and $6$ are given in Sec. V,
and we give concluding remarks in Sec. VI.
\section{Fluid Membranes}
In this paper, we use the term membrane to denote a thin film of one material
that separates two similar materials (bilayer membrane). We focus on fluid
membranes and their possible degrees of orientational order. 
In fluid memebranes,
there is no in-plane shear modulus and the only in-plane deformations are
compressions/expansions. An $n$-atic membrane, although fluid, will sustain a
local orientational order defined modulo rotations by $2\pi/n$, 
which is described by a local order parameter $\psi = \exp[in\Theta]$.
\par
Although fluid membranes can be composed of many different types of chemical
and molecular species, their behavior (shapes, fluctuations, thermodynamics) 
can be understood from a unified point of view that considers their free
energy of deformation.
If the membrane were constrained to lie in a plane, the only relevant energy 
would be the compression of the molecules, that is, change of the average area
per molecule. This is analogous to sound waves in a 3-dimensional fluid; there
is no low-frequency reponse of the system to shear.
However, since the membrane can also deform in the normal direction,
there is an additional set of modes describing the conformations of the film.
These out-of-plane deformations are known as bending or curvature modes and 
the free energy associated with such modes is known as the curvature free
energy. 
While a general deformation of the membrane involves changes in both 
volume and curvature, we shall see that the lowest energy deformations
usually involve only curvature. 
In most systems, changing the average 
volume of the membrane is a higher energy process and hence is less important
when effects involving the thermal behavior of the membrane are considered.
For a membrane with finite thickness, we denote as pure curvature deformations
those perturbations of the membrane that do not change the overall membrane
volume, but where there may be local stretching and compression of different
parts of the film. The location of the interface within a membrane can be 
chosen to lowest order in the curvature energy so that the surface which
defines the interface undergoes no stretching or compression (neutral surface).
For long-wavelength curvature deformations whose wave lengths are much larger
than the membrane thickness, the exact position of the neutral surface within 
the membrane is not crucial.
\par
The membrane is characterized by the area per molecule, $\Sigma$, 
its thickness, $\lambda$, and its curvature. For simplicity, 
we assume that the equation of state of the membrane
determines the thickness as a function of the area per molecule. A simple
example is the case of an incompressible molecule where the product of
$\lambda\Sigma$
is constrained to equal the molecular volume so that $\lambda \sim 1/\Sigma$.
We thus take the curved membrane to be characterized only by the area
per molecule and its curvature.
Consider a locally flat, isolated interface. Saturation occurs when the 
free energy achieves a minimum with respect to $\Sigma$;
$\partial f_{0}/\partial\Sigma = 0$ where $f_{0}$ is the free energy per 
molecule for a flat layer and $\Sigma$ is the area per molecule.
The free energy per molecule is minimized when $\Sigma=\Sigma_{0}$, 
the optimal value of the
area per molecule which arises from a balance of terms such as the entropy
and the tension terms or attractions. The entropy favors a larger area per 
molecule because of the greater number of center-of-mass positions and
chain conformations, while the tension terms and attractions favor
a small value of $\Sigma$ to reduce the contact of the 
hydrocarbon chains with the water.
There can be deviations in the area per molecule from this optimal value
and the energy cost of such a compression or expansion is
\begin{equation}
\Delta f_{0} =\frac{1}{2}f_{0}''(\Sigma-\Sigma_{0})^{2}
\end{equation}
where the primes signify a derivative with respect to $\Sigma$.
However, these deformations are typically of higher energy than the curvature
deformations. A membrane can change its shape
or size with a much lower free energy cost than that required to compress
or expand it. It is important to remember, therefore, that for insoluble 
amphiphiles it is the saturation of the membrane and the minimization of 
the area per molecule that permits the usual surface-tension term
to be neglected; $\partial f / \partial\Sigma = 0$.
The surface tension is no longer relevant since the molecules
adjust their area to optimize the free energy and it is the 
curvature energy that determines the properties of the membrane.
\par
To construct the effective free energy for fluid membranes, it is convenient
to introduce general curvilinear coordinates ${\bf u}= \{ u^{1},u^{2} \}$
and the metric structure. Defining locally a system of coordinates 
${\bf u}$ on the membrane and denoting by ${\bf X}({\bf u})$ the position of 
the point ${\bf u}$ in bulk $d$-dimensional Euclidean space, the metric tensor 
$g_{\alpha\beta}({\bf u})$ induced by the embedding is
\begin{equation}
g_{\alpha\beta}({\bf u}) = {\bf t}_{\alpha}({\bf u}) \cdot 
{\bf t}_{\beta}({\bf u})\;  ; \;\;
{\bf t}_{\alpha}=\frac{\partial{\bf X}({\bf u})}{\partial u^{\alpha}},
\end{equation}
where ${\bf t}_{\alpha}$ is a tangent vector to the surface, and the element
of area is
\begin{equation}
d^{2}s = d^{2}u\sqrt{g} \; ; \;\; g=\det(g_{\alpha\beta}).
\end{equation}
The extrinsic curvature tensor ${\bf K}_{\alpha\beta}$ is defined by
\begin{equation}
{\bf K}_{\alpha\beta}({\bf u}) = D_{\alpha}D_{\beta}{\bf X}({\bf u}),
\end{equation}
where $D_{\alpha}$ is a covariant derivative with respects to 
the metric $g_{\alpha\beta}$. In general, ${\bf K}_{\alpha\beta}$ is 
normal to the surface [3]. Therefore, in the 
particular case of a surface in $R^{3}$, ${\bf K}_{\alpha\beta}$ is 
proportional to the unit normal vector ${\bf N}$ to the surface and it is 
written as
\begin{equation}
{\bf K}_{\alpha\beta} = K_{\alpha\beta}{\bf N},
\end{equation}
where $K_{\alpha\beta}$ is a symmetric tensor.
We can dicuss the curvature energy using symmetry considerations.
The free energy must be only a function of the field ${\bf X}({\bf u})$, 
invariant under displacements and rotations in $R^{d}$, 
and reparametrization-invariant.
Expanding in local terms involving more and more derivatives and keeping only
the terms relevant by naive power counting, the most general form of the 
curvature energy ${\cal H}_{\rm c}$ up to quadratic order in the
curvatures can be written in terms of the mean and Gaussian curvatures and
has only two terms
\begin{equation}
{\cal H}_{\rm c}=\frac{1}{2}\kappa\int d^{2}u\sqrt{g}(H-H_{0})^{2} + 
             \frac{1}{2}\bar{\kappa}\int d^{2}u\sqrt{g} K.
\end{equation}
Here, $H$ is the mean curvature, $\frac{1}{2} K_{\alpha}^{\alpha}$, and 
$K$ is the Gaussian curvature, $\det K_{\alpha}^{\beta}$.
This form for the curvature energy was discussed by Helfrich and Canham
and states
that the mean curvature that minimizes the energy has a value $H_{0}$,
termed the spontaneous curvature of the membrane \cite{Helfrich-73}. 
The energy cost of deviating
from the spontaneous curvature is the bending or curvature modulus, $\kappa$.
The parameter $\bar{\kappa}$, known as the saddle-splay modulus, measures
the energy cost of saddle-like deformation.
The spontaneous curvature describes the tendency of the bilayer membrane to
bend. It is viewed to arise from the fact that the two layers of the bilayer
do not have areas per molecule which are exactly equal, nor are their 
curvatures equal and opposite. This can result in a vesicle that is more 
stable than a flat, lamellar bilayer since it may be that the favorably curved
outer layer has a smaller value of the area per molecule and 
hence more molecules in it than the inner layer. The bending moduli,
$\kappa$ and $\bar{\kappa}$, arise from the elastic constants determined by
the head-head and tail-tail interactions. It is expected that these moduli 
are senstively dependent on the surfactant chain length but only weakly
dependent on the head-head interaction strength. The parameters $H_{0}$,
$\kappa$, and $\bar{\kappa}$ can be derived from a simple microscopic
model that incorporates both the change in the area per molecule and
the curvature. We note that a stable membrane will always have $\kappa > 0$.
However, the sign of $\bar{\kappa}$ can be either positive or negative.
Membranes that prefer isotropic shapes where the Gaussian curvature
$K > 0$ such as spheres or planes will have $\bar{\kappa} > 0$, while
membranes that prefer saddle shapes where the Gaussian curvature $K < 0$
will have $\bar{\kappa} < 0$. One can show that the requirement of a
positive-definite quadratic term implies that membranes are only stable
if $\kappa-2\bar{\kappa} > 0$; otherwise higher order curvature terms are
needed to stabilize the system. 
We consider the spherical membrane whose topology is fixed. Thus the 
Gaussian curvature energy gives constant contribution by Gauss-Bonnet
theorem $\int\sqrt{g} K = 2\pi\chi$.
\section{Orientational Order}
In a fluid membrane, molecules can flow freely to adapt themselves 
to any particular shape of the surface. 
If correlations among the molecular positions of the molecules
forming the membrane exist, the molecules may exhibit in-plane crystalline
order and form a kind of two-dimensional solid. On the other hand, the
molecules may exhibit a weaker order in which the orientations are correlated
at long distance scale. Orientational order means that to each point on the
membrane is associated a preferred direction whithin the tangent plane of the
membrane.
For example at high temperature, the stable phase of the membrane is 
generally the liquid phase with no translational or orientational order if we 
are considering the hexatic order,
or the smectic-A phase in which molecular axes are 
normal to the surface for the vector order. 
At lower temperature, the membrane can condense into an 
hexatic phase in which there is quasi-long-range 6-fold bond-angle order or
into a smectic-C phase in which molecules tilt relative to the surface normal.
To describe Sm-C (vector) and/or hexatic order, we introduce at each point 
${\bf X}({\bf u})$
on the membrane a unit vector ${\bf m}({\bf u})$ in the tangent plane of the 
membrane. For Sm-C order, ${\bf m}({\bf u})$ is a true vector, invariant under
rotations of $2p\pi$ (p is an integer) about the unit surface normal
${\bf N}({\bf u})$ erected at ${\bf X}({\bf u})$. For hexatic order, 
rotations of ${\bf m}({\bf u})$ by $2p\pi/6$ about ${\bf N}({\bf u})$ 
lead to physically equivalent states. 
More generally, we consider ``$n$-atic'' order in which rotations of 
${\bf m}({\bf u})$ by $2p\pi/n$ produce physically equivalent states. A 
two-dimensional nematic with an in-plane symmetric traceless tensor order 
parameter is an example of a $2$-atic. Although we know of no physical
realizations of other $n$-atics yet, we find it instructive to consider how the
development of such order affects morphological changes in spherical vesicles.
\par
To describe tangent-plane $n$-atic order, we introduce orthonormal unit 
vectors ${\bf e}_{1}$ and ${\bf e}_{2}$ at each point on the membrane.
${\bf e}_{1}({\bf u})\cdot {\bf m}({\bf u}) = \cos\Theta ({\bf u})$ defines
a local angle $\Theta ({\bf u})$. $n$-atic order is then described by the
order parameter $\psi ({\bf u}) = \exp[in\Theta({\bf u})]$,
which can be related to the $n$-th rank symmetric traceless tensor 
constructed from the unit vector ${\bf m}$. The $n$-th rank symmetric 
traceless tensors are the $n$-th rank spherical tensors. They are listed
in Table~\ref{n-tensor} for $n=1,2,3,4,$ and 6.
In general, there are $2^{n}$ components in the $n$-th rank spherical
tensor ${\bf Q}^{(n)}_{i_{1} i_{2} \cdots i_{n}}$ since $i_{1},i_{2},\cdots,
i_{n}$ can be either $1$ or $2$. By permutational symmetry, there are only 
$(n+1)$ possible independent components. However, there are $(n-1)$ additional
traceless conditions. Hence, there are only $2$ independent components in
${\bf Q}^{(n)}_{i_{1} i_{2} \cdots i_{n}}$. The linear combination of these
$n$-th rank spherical tensors
\begin{equation}
\sum^{1,2}_{i_{1},i_{2},\cdots,i_{n}} i^{k} 
{\bf Q}^{(n)}_{i_{1} i_{2} \cdots i_{n}},
\end{equation}
where $k$ is the number of $2's$ in $(i_{1},i_{2},\cdots,i_{n}),$ becomes
$(m_{1}+i m_{2})^{n}$ where $m_{1}={\bf e}_{1}\cdot{\bf m}=\cos\Theta$ and
$m_{2}={\bf e}_{2}\cdot{\bf m}=\sin\Theta$. Thus $n$-atic order parameter is
described by
\begin{equation}
\psi=\sum^{1,2}_{i_{1},i_{2},\cdots,i_{n}} i^{k} 
{\bf Q}^{(n)}_{i_{1} i_{2} \cdots i_{n}}
    =e^{in\Theta}.
\end{equation}
Note that since $\Theta({\bf u})$ depends on the choice of orthonormal vectors
${\bf e}_{1}$ and ${\bf e}_{2}$, the order parameter $\psi({\bf u})$ does
as well. This means that any spatial derivatives in a phenomenological
Hamiltonian for $\psi$ must be covariant derivatives.
\par
We now describe the Hamiltonian of the membrane with orientational order
in a reparametrization-invariant way. For a flat membrane the vector
Hamiltonian corresponds to the usual XY model
\begin{equation}
{\cal H}=\frac{1}{2} K_{n}\int d^{2}x
\partial_{\alpha}{\bf m}\cdot\partial_{\alpha}{\bf m}.
\end{equation}
Using the fact that any spatial derivatives must be covariant derivatives
and 
\begin{equation}
D_{\alpha}{\bf m} = (D_{\alpha}m^{\beta}){\bf t}_{\beta}
\end{equation}
is the tangential component of $\partial_{\alpha}{\bf m}$, one can show that 
the only possible free energy term for ${\bf m}$ which respects this $Z_{1}$
symmetry is
\begin{equation}
{\cal H}=\frac{1}{2} K_{1}\int d^{2}u\sqrt{g}
D_{\mu}m^{\alpha}D^{\mu}m_{\alpha}.
\end{equation}
The dimensionless coupling constant $K_{1}$ is the vector-order stiffness and
measures the strength of the coupling between the orientations of neighboring 
vectors. 
We then generalize this XY-like Hamiltonian for the vector order to
the $n$-atic Hamiltonian using the $n$-th rank spherical tensors introduced
in Table~\ref{n-tensor}.
In general, the $n$-atic Hamiltonian can be written as
\begin{equation}
{\cal H}=\frac{1}{2} K_{n} \int d^{2}u\sqrt{g}
\frac{D_{\mu}{{\bf Q}^{(n)}}^{\alpha_{1}\cdots\alpha_{n}}
D^{\mu}{{\bf Q}^{(n)}}_{\alpha_{1}\cdots\alpha_{n}}}
{{{\bf Q}^{(n)}}^{\alpha_{1}\cdots\alpha_{n}}
{{\bf Q}^{(n)}}_{\alpha_{1}\cdots\alpha_{n}}},
\end{equation}
where $K_{n}$ is the $n$-atic rigidity.
Using the spherical tensors in Table 1, we find
\begin{equation}
{{\bf Q}^{(n)}}^{\alpha_{1}\cdots\alpha_{n}}
{{\bf Q}^{(n)}}_{\alpha_{1}\cdots\alpha_{n}} = \frac{1}{2(n-1)},
\end{equation}
and
\begin{equation}
D_{\mu}{{\bf Q}^{(n)}}^{\alpha_{1}\cdots\alpha_{n}}
D^{\mu}{{\bf Q}^{(n)}}_{\alpha_{1}\cdots\alpha_{n}} =
\frac{n^{2}}{2(n-1)} D_{\mu}m^{\alpha}D^{\mu}m_{\alpha},
\end{equation}
for $n > 1$.
Hence, the Hamiltonian for the $n$-atic order becomes
\begin{equation}
{\cal H}_{n} = \frac{1}{2} n^{2}K_{n} \int d^{2}u\sqrt{g}
D_{\mu}m^{\alpha}D^{\mu}m_{\alpha}.
\end{equation}
In this form of ${\cal H}_{n}$, we neglected all terms that are irrelevant
at large distance by power counting. Other terms such as $m_{\alpha}
K^{\alpha}_{\beta}K^{\beta\gamma}m_{\gamma}$ which couple ${\bf m}$ to
the principal directions of curvature of the membrane 
are not invariant under the global rotation by $2\pi/n$ of
${\bf m}$, and are irrelevant at large distance. Thus, this free energy has 
a full $O(2)$ rotational symmetry. This is similar to the fact a 2-dimensional
crystal with hexagonal or triangular structure has isotropic elastic 
properties at large distance scale. For $n$-atics with $n \geq 3$, 
there is only one elastic constant $K_{n}$. 
For $n=1$ or $n=2$, there are in general two
elastic constants. For simplicity, we will consider the single elastic constant
approximation for all $n$-atics.
\par
To describe this Hamiltonian in terms of a local angular order
parameter $\Theta$,
at each point we introduce two orthonormal vectors ${\bf e}_{a}({\bf u}) 
(a=1,2)$
tangent to the membrane. This is equivalent to introduce a 2-{\it bein}
$e^{i}_{a}({\bf u})$ compatible with the induced metric 
$g_{\alpha\beta}({\bf u})$.  In components,
\begin{equation}
{\bf e}_{a} = e^{\alpha}_{a} {\bf t}_{\alpha}
\end{equation}
and the orthonormality ${\bf e}_{a}\cdot{\bf e}_{b} = \delta_{ab}$ implies
\begin{equation}
e^{\alpha}_{a}e^{\beta}_{b}g_{\alpha\beta} = \delta_{ab}\; ; \;\; 
e^{\alpha}_{a}e^{\beta}_{b}\delta^{ab} = g^{\alpha\beta}.
\end{equation}
The angular order parameter is frustrated by the rotation of tangent vectors
that occurs under parallel transport on a curved surface. The amount of
frustration is given by the gauge field $A_{\alpha}$, {\it i.e.}
the covariant derivative of ${\bf e}_{a}$ in direction $\alpha$ defines the
gauge field $A_{\alpha}$. 
Under parallel transport in direction $du^{\alpha}$, each ${\bf e}_{a}$ is 
rotated by an angle $A_{\alpha} du^{\alpha}$.
Thus the gauge field $A_{\alpha}$ is defined by
\begin{equation}
D_{\alpha} {\bf e}_{a} = -A_{\alpha}\varepsilon_{ab}{\bf e}_{b},
\end{equation}
where $\varepsilon_{ab}$ is the antisymmetric tensor with
$\varepsilon_{12}=-\varepsilon_{21}=1$
and $A_{\alpha}\varepsilon_{ab}$ is called the spin-connection and describes
how the basis vector ${\bf e}_{a}$ rotates under parallel transport according
to the Gaussian curvature $K$ of the surface.
In fact, $A_{\alpha}$ is related to $K$.
The curl of the gauge field $A_{\alpha}$ is the Gaussian curvature;
\begin{equation}
\gamma^{\alpha\beta}D_{\alpha}A_{\beta} = K,
\label{curl-spin}
\end{equation}
where $\gamma^{\alpha\beta}$ is the antisymmetric tensor defined via
\begin{eqnarray}
\gamma_{\alpha\beta} & = & {\bf N}\cdot ({\bf t}_{\alpha}
\times{\bf t}_{\beta}) =
\sqrt{g}\varepsilon_{\alpha\beta},  \nonumber  \\
\gamma^{\alpha\beta} & = & g^{\alpha\alpha'}g^{\beta\beta'}
\gamma_{\alpha'\beta'},
\label{antisym}
\end{eqnarray}
The covariant derivative of ${\bf m}=\cos\Theta {\bf e}_{1} + 
\sin\Theta {\bf e}_{2}$ writes
\begin{eqnarray}
D_{\alpha} {\bf m} & = & (D_{\alpha} m_{a}) {\bf e}_{a} +
                         m_{a} (D_{\alpha}{\bf e}_{a})  \nonumber  \\
   & = & (D_{\alpha} m_{a}) {\bf e}_{a} -
         m_{a} A_{\alpha}\varepsilon_{ab} {\bf e}_{b}  \nonumber  \\
   & = & (D_{\alpha}\Theta) (-\sin\Theta {\bf e}_{1} + \cos\Theta {\bf e}_{2})
  \nonumber  \\
   &   & - A_{\alpha} (\cos\Theta {\bf e}_{2} - \sin\Theta {\bf e}_{1}) 
  \nonumber \\
   & = & (D_{\alpha}\Theta - A_{\alpha}) {\bf m}_{\perp},
\end{eqnarray}
where ${\bf m}_{\perp} = -\sin\Theta {\bf e}_{1} + \cos\Theta {\bf e}_{2}$
satisfying ${\bf m}\cdot{\bf m}_{\perp} = 0$.
Then the $n$-atic Hamiltonian writes
\begin{equation}
{\cal H}_{n}  = \frac{1}{2}n^{2}K_{n} \int d^{2}u \sqrt{g}
g^{\alpha\beta}(\partial_{\alpha}\Theta-A_{\alpha}) 
(\partial_{\beta}\Theta-A_{\beta}).
\end{equation}
This form of the free energy is invariant under local transformations
$\Theta({\bf u}) \rightarrow \Theta({\bf u}) + \Lambda({\bf u})$, 
$A_{\alpha}({\bf u}) \rightarrow A_{\alpha}({\bf u}) + 
\partial_{\alpha}\Lambda({\bf u})$. This gauge invariance
corresponds to a local rotation of the reference frame ${\bf e}_{a}$.
In terms of the complex order parameter $\psi$, the Hamiltonian for
$n$-atic order becomes
\begin{equation}
{\cal H}_{n}=\frac{1}{2}K_{n}\int d^{2}u\sqrt{g}g^{\alpha\beta}
D_{\alpha}\psi (D_{\beta}\psi)^{*}
\end{equation}
where $D_{\alpha}\psi=(\partial_{\alpha}-inA_{\alpha})\psi$. 
In this description, all 
$n$-atics have the same long-wavelength elastic energy. Their properties
can differ however, because their topological excitations are characterized
by different winding numbers.
Since we assume that in the disordered state the membrane forms a spherical 
shape and there is no topological deformation of the membrane shape, order in
$\psi$ is necessarily accompanied by the topological excitations called
vortices. As we can see in Ref. \cite{Spivak-79}, 
the total winding vorticity of a vector
field on a closed surface with $h$ handles must be $2(1-h)$. Thus, a vector
field on the surface of a sphere $(h=0)$ has total vorticity 2. The minimum
winding number disclination for an $n$-atic is $1/n$. The energy of an 
individual disclination on both flat and curved surfaces is proportional to the
square of its winding number. It is therefore always favorable to form
disclinations with the lowest possible winding number. In addition, 
disclinations with the same sign repel each other. These considerations imply 
that the ground state of a sphere with surface $n$-atic order will have $2n$
maximally separated disclinations of winding number $1/n$.
\section{Existence of the KT transition}
In this section, we present that in a certain region in the parameter
space $(\kappa,K_{n})$ there is an effective KT transition.
Ovrut and Thomas have discussed the structure of the KT transition
of a vortex-monopole Coulomb gas on a rigid sphere and have shown it
is the same as in the planar case. They have shown the KT transition
temperature on a rigid sphere is the same as that on the Euclidean plane,
{\it i.e.} $T^{\rm KT}_{\rm sphere} = T^{\rm KT}_{\rm plane}
= \pi K_{A}/2$, where $K_{A}$ is the square of the vortex charge in 
their discription \cite{Ovrut-91}.
Recently, we have extended the study of the structure of the KT transition
to the hexatic Hamiltonian on the fluctuating spherical surfaces 
and have shown the hexatic Hamiltonian undergoes the effective KT transition
at the same tenperature as in the planar case \cite{Park-96}.
Following the procedure in Ref. \cite{Park-96},
we generalize a study of the KT transition to the $n$-atic Hamiltonian
and present the region in the parameter space $(\kappa,K_{n})$ where
the KT transition exists.
\par 
From the previous sections, the $n$-atic Hamiltonian on the fluctuating
surfaces writes ${\cal H}={\cal H}_{c}+{\cal H}_{n}$;
\begin{eqnarray}
{\cal H}_{\kappa} & = & \frac{1}{2} \int d^{2}u\sqrt{g} (H-H_{0})^{2} \\
{\cal H}_{n} & = & \frac{1}{2}n^{2}K_{n} \int d^{2}u\sqrt{g} g^{\alpha\beta}
(\partial_{\alpha}\Theta - A_{\alpha})(\partial_{\beta}\Theta - A_{\beta}).
\end{eqnarray}
We investigate the effect of thermal shape fluctuations of a genus zero
surface on the KT transition in the low temperature limit $\beta\kappa \gg 1$.
In this limit, we can parametrize the surface by its radius vector
as a function of standard polar coordinates ${\bf u}=(\theta,\phi)
\equiv \Omega$;
\begin{equation}
{\bf R}(\Omega) = R(1+\rho(\Omega)){\bf e}_{r},
\end{equation}
where ${\bf e}_{r}$ is the radial unit vector and $\rho(\Omega)$ measures
deviation from sphericity with radius $R$.
This parametrization is a ``normal gauge''.
To make the equations simple, we map this parametrization onto the
unit sphere parametrization with $R=1$.
Since we are interested in the limit $\beta\kappa \gg 1$ and
in this limit ${\cal H}_{c}$ dominates, we first minimize
${\cal H}_{c}$ over the shape fluctuation field $\rho$ which gives
$\rho(\Omega) = 0$ and
then we minimize ${\cal H}_{n}$ over $\Theta$ with $\rho(\Omega)=0$ and find
\begin{equation}
\left. \frac{\delta{\cal H}^{0}_{n}}{\delta\Theta(\Omega)} 
\right|_{\Theta=\Theta^{0}} = \frac{1}{\sqrt{g^{0}}} \partial_{b}
{g^{0}}^{ab} (\partial_{a}\Theta^{0}-A^{0}_{a}) = 0,
\end{equation}
where the superscript$~^{0}$ stands for the rigid sphere with 
$\rho(\Omega)=0$.
In Ref.~\cite{Lubensky-92}, Lubensky and Prost show that in the ground state
$2n$ disclinations of strength $2\pi/n$ are arranged at the vertices of 
polyhedron inscribed in the sphere.
A disclination at ${\bf u} = {\bf u}_{i}$ with 
strength $q_{i}$ gives rise to a singular contribution $\Theta^{\rm sing}_{0}$
to $\Theta^{0}$ satisfying 
\begin{equation}
\oint_{\Gamma} du^{\alpha}\partial_{\alpha}\Theta^{\rm sing}_{0} = q_{i},
\end{equation}
where $\Gamma$ is a contour enclosing ${\bf u}_{i}$.
Thus, in general $\partial_{\alpha}\Theta^{0} = \partial_{\alpha}\Theta'_{0}
+v^{0}_{\alpha}$ where $\Theta'_{0}$ is nonsingular, $v^{0}_{\alpha}=
\partial_{\alpha}\Theta^{\rm sing}_{0}$, and
\begin{equation}
\gamma^{\alpha\beta}D_{\alpha}v^{0}_{\beta} = s^{0}(\Omega),
\end{equation}
where
\begin{equation}
s^{0}(\Omega) = \frac{2\pi}{n} \sum^{2n}_{i=1} 
\delta(\Omega-\Omega_{i}),
\end{equation}
where is the disclination density in the ground state and
$\Omega_{i}$'s are the coordinates of the vertices of icosahedron.
Since $(\partial_{\alpha}\Theta^{0}-A_{\alpha}^{0})$ satisfies 
$D^{\alpha}(\partial_{\alpha}\Theta^{0}-A_{\alpha}^{0}) = 0$,
it is divergence-less and purely transverse.
Accordingly $(\partial_{\alpha}\Theta^{0}-A_{\alpha}^{0})$ can be written
in terms of the curl of scalar fields and by applying the operator 
$\gamma^{\beta\alpha} D_{\beta}$ to 
$(\partial_{\alpha}\Theta^{0}-A^{0}_{\alpha})$, we find
these scalar fields to be related to the Gaussian curvature $K_{0}$ of
the rigid sphere and the ground state disclination density on the rigid
sphere,
\begin{eqnarray}
\gamma^{\beta\alpha} D_{\beta}
(\partial_{\alpha}\Theta^{0}-A^{0}_{\alpha}) & = & 
    \gamma^{\beta\alpha}D_{\beta}v^{0}_{\alpha} - 
    \gamma^{\beta\alpha}D_{\beta}A^{0}_{\alpha}  \nonumber  \\
   & = & s^{0} - K_{0},
\end{eqnarray}
where $K_{0}$ is a Gaussian curvature of the rigid sphere and
$s^{0}$ is the disclination density in the ground state.
\par
Now taking into account the bond angle fluctuations around $\Theta^{0}$ and
the shape fluctuations around the sphere, 
\begin{equation}
\Theta = \Theta^{0} + \tilde{\Theta}, \;\;
A_{\alpha} = A^{0}_{\alpha} + \delta A_{\alpha},
\end{equation}
the full
Hamiltonian writes ${\cal H}={\cal H}_{0}+\delta{\cal H}$
\begin{eqnarray}
{\cal H}_{0} & = & \frac{1}{2}n^{2}K_{n} \int d\Omega 
(\partial^{\alpha}\Theta^{0}-{A^{0}}^{\alpha})(\partial_{\alpha}\Theta^{0}
-A^{0}_{\alpha})
\nonumber  \\
\delta{\cal H} & = & \frac{1}{2}\kappa \int d\Omega \left( (\nabla^{2}+2)\rho
\right)^{2}
\nonumber  \\
   &   & + \frac{1}{2}n^{2}K_{n} \int d\Omega
(\partial^{\alpha}\tilde{\Theta}-\delta A^{\alpha})
(\partial_{\alpha}\tilde{\Theta}-\delta A_{\alpha}) + {\cal O}(\rho^{3}).
\end{eqnarray}
The angle fluctuation field $\tilde{\Theta}(\Omega)$ can also have 
disclinations of strength $q=2\pi(k/n)$ where $k$ is an integer,
due to the thermal fluctuation \cite{Lubensky-95}.
Thus, $\partial_{\alpha}\tilde{\Theta}$ can be decomposed into singular
and nonsingular parts;
$\partial_{\alpha}\tilde{\Theta} = \partial_{\alpha}\Theta^{\parallel} + 
v_{\alpha}$ where $\Theta^{\parallel}$ is
nonsingular, $v_{\alpha} = \partial_{\alpha}\tilde{\Theta}^{\rm sing}$ and
\begin{equation}
\gamma^{\alpha\beta}D_{\alpha} v_{\beta} = s(\Omega),\;\;
s(\Omega) = \sum_{i} q_{i} \delta(\Omega - \Omega_{i}),
\label{eq:vortex}
\end{equation}
where $s(\Omega)$ is the thermally-excited disclination density
with disclinations of strength $q_{i}$ at $\Omega_{i}$.
The vector $v_{\alpha}$ can always be chosen so that it is purely transverse,
{\it i.e.} $D_{\alpha}v^{\alpha} = 0$. 
In the hexatic Hamiltonian, $\partial_{\alpha}\tilde{\Theta}$ always occurs 
in the combination
$(\partial_{\alpha}\tilde{\Theta} - \delta A_{\alpha})$. 
The spin-connection $\delta A_{\alpha}$ can and will
in general have both a longitudinal and a transverse component.
However, one can always redefine $\Theta^{\parallel}$ to include the 
longitudinal part of $\delta A_{\alpha}$. This amounts to choosing 
locally rotated orthonormal vectors
${\bf e}_{1}({\bf u})$ and ${\bf e}_{2}({\bf u})$ so that 
$D_{\alpha} \delta A^{\alpha} = 0$.
Thus we may take both $v_{\alpha}$ and $\delta A_{\alpha}$ to be transverse and
the hexatic Hamiltonian
\begin{eqnarray}
   &   & \frac{1}{2}n^{2}K_{n} \int d\Omega 
(\partial^{\alpha} \Theta^{\parallel}+v^{\alpha} -\delta A^{\alpha})
(\partial_{\alpha} \Theta^{\parallel}+v_{\alpha} -\delta A_{\alpha})
\nonumber  \\
   & = & {\cal H}_{\parallel} + {\cal H}_{\perp},
\end{eqnarray}
can be decomposed into a regular longitudinal part,
\begin{equation}
{\cal H}_{\parallel} =  \frac{1}{2}n^{2}K_{n}\int d\Omega
\partial^{\alpha}\Theta^{\parallel}\partial_{\alpha}\Theta^{\parallel},
\end{equation}
and a transverse part,
\begin{equation}
{\cal H}_{\perp} =  \frac{1}{2}n^{2}K_{n}\int d\Omega
(v^{\alpha} - \delta A^{\alpha})(v_{\alpha} - \delta A_{\alpha}),
\label{eq:h-perp}
\end{equation}
where the cross term $\int d\Omega 
(v_{\alpha}-\delta A_{\alpha}) \partial^{\alpha}\Theta^{\parallel}$ 
is dropped since
$D^{\alpha}(v_{\alpha}-A_{\alpha}) = 0$.
\par
It costs an energy $\epsilon_{c}(k)$ to create the core of a disclination
of strength $k$. (We assume for the moment that the core energies of the
positive and negative disclinations are the same.  See, however, Refs.
\cite{jp-lub,jp-lub3,Deem-96}.) 
Thus, partition sums should be weighted by a factor
$y_{k} = e^{-\beta\epsilon_{c}(k)}$ for each disclination of strength
$k$. Since $\epsilon_{c}(k) \sim k^{2}$, we may at low temperature
restrict our attention to configurations in which only configurations
of strength $\pm 1$ appear. Let $N_{\pm}$ be the number of disclinations
of strength $\pm 1$ and let ${\bf u}_{i^{\pm}}$ be the coordinate of the 
core of
the disclination with strength $\pm 1$ labeled by $i$. The hexatic membrane
partition function can then be written as
\begin{equation}
{\cal Z} (\kappa, K_{A}, y) = {\rm Tr}_{\rm v} y^{N}
\int{\cal D}{\bf R}\int{\cal D}\Theta^{\parallel}
e^{-\beta{\cal H}_{\kappa}}e^{-\beta({\cal H}_{\parallel}+
{\cal H}_{\perp})},
\label{eq:partition}
\end{equation}
where $y=y_{1}$, and $N = N_{+} + N_{-}$.
${\cal H}_{\perp}$ depends on all of the disclination coordinates
$\Omega_{\nu^{\pm}}$ where $\nu^{\pm}=1,2,\cdots,N_{\pm},$ and
${\rm Tr}_{\rm v}$ is the sum over all possible disclination
distribution with the topological 
constraint \cite{Spivak-79};
\begin{equation}
{\rm Tr}_{\rm v} = \sum_{N_{+},N_{-}} \frac{\delta_{N_{+},N_{-}}}
{N_{+}!N_{-}!} \prod_{\nu^{+}}\int \frac{d\Omega_{\nu^{+}}}{a^{2}}
\prod_{\nu^{-}}\int \frac{d\Omega_{\nu^{-}}}{a^{2}},
\end{equation}
where $a^{2}$ is a molecular solid angle. 
The Kronecker factor $\delta_{N_{+},N_{-}}$
in ${\rm Tr}_{\rm v}$ imposes the topological constraint that the total 
disclination strength on a sphere is 2 since with $N_{+}=N_{-}$ we have $2n$
ground state disclinations with the strength $1/n$ 
giving the total disclination
strength $2n \times (1/n) =2$.
\par
The hexatic model of Eq.~(\ref{eq:partition}) can easily be converted to a 
Coulomb gas model using
\begin{equation}
\gamma^{\alpha\beta}D_{\alpha}(v_{\beta}-\delta A_{\beta}) = 
s - \delta K \equiv {\cal C},
\label{eq:curl-vel}
\end{equation}
which follows from Eq.~(\ref{curl-spin}) and Eq.~(\ref{eq:vortex}) where
$\delta K$ is the deviation of the Gaussian curvature from the rigid sphere.
The quantity ${\cal C} = s-\delta K$ is a ``charge'' density 
with contributions arising 
both from disclinations and from Gaussian curvature.
Equation (\ref{eq:curl-vel}) implies
\begin{equation}
v_{\alpha}-\delta A_{\alpha} = -{\gamma_{\alpha}}^{\beta}D_{\beta} 
\frac{1}{\Delta}{\cal C},
\label{eq:va-Aa}
\end{equation}
where we used $\gamma_{\alpha\lambda}D^{\lambda}\gamma^{\alpha\beta}D_{\alpha}
= - \Delta$ and 
$\Delta=D^{\alpha}D_{\alpha} = (1/\sqrt{g})\partial_{\alpha}\sqrt{g}
g^{\alpha\beta}\partial_{\beta}$ 
is the Laplacian on a surface with metric tensor 
$g_{\alpha\beta}$ acting on a scalar. 
Recall [Eq.~(\ref{antisym})] that ${\gamma_{\alpha}}^{\beta}$ rotates
a vector by $\pi/2$ so that $(v_{\alpha}-\delta A_{\alpha})$ is perpendicular 
to
$D_{\beta}(-\Delta)^{-1}{\cal C}$ and is thus manifestly transverse.
Using Eq.~(\ref{eq:va-Aa}) in Eq.~(\ref{eq:h-perp}), we obtain
\begin{equation}
{\cal Z} = {\rm Tr}_{\rm v} y^{N}\int{\cal D}{\bf R}
\int{\cal D}\Theta^{\parallel} e^{-\beta
{\cal H}_{\kappa} -\beta{\cal H}_{\parallel} -\beta{\cal H}_{\rm c}},
\end{equation}
where
\begin{eqnarray}
{\cal H}_{\rm C} & = & \frac{1}{2}n^{2}K_{n} \int d\Omega 
\frac{\gamma_{\alpha}^{\beta}D_{\beta}}{\Delta} {\cal C}(\Omega)
\frac{\gamma^{\alpha\lambda}D_{\lambda}}{\Delta}{\cal C}(\Omega)  \nonumber  \\
   & = & \frac{1}{2}n^{2}K_{n} \int d\Omega d\Omega'
{\cal C}(\Omega) \left( \frac{\gamma_{\alpha}^{\beta}D_{\beta}
\gamma^{\alpha\lambda}D_{\lambda}}{\Delta^{2}} \delta(\Omega-\Omega')
\right) {\cal C}(\Omega')   \nonumber  \\
   & = & \frac{1}{2}n^{2}K_{n} \int d\Omega d\Omega'
{\cal C}(\Omega) \left( -\frac{1}{\Delta} \delta(\Omega-\Omega') \right)
{\cal C}(\Omega'),
\label{coulomb-rr}
\end{eqnarray}
is the Coulomb Hamiltonian associated with the charge ${\cal C}$.
Since the longitudinal variable $\Theta^{\parallel}$ appears only 
quadratically in
${\cal H}_{\parallel}$, the trace over $\Theta^{\parallel}$ can be done 
directly giving the Liouville action \cite{Polyakov-81} 
arising from the conformal anomaly;
\begin{equation}
\int {\cal D}\Theta^{\parallel} e^{-\beta{\cal H}_{\parallel}} = 
e^{-\beta{\cal H}_{\rm L}},
\end{equation}
where
\begin{equation}
\beta{\cal H}_{\rm L} = \frac{1}{8\pi a^{2}}\int d\Omega -
\frac{1}{24\pi}\int d\Omega d\Omega' K(\Omega)
\left(-\frac{1}{\Delta} \delta(\Omega-\Omega')\right) K(\Omega').
\end{equation}
\par
The Coulomb gas partition function can thus be written
\begin{equation}
{\cal Z} = {\rm Tr}_{\rm v} y^{N}\int{\cal D}{\bf R} 
e^{-\beta{\cal H}_{\kappa}
-\beta{\cal H}_{\rm L} -\beta{\cal H}_{\rm C}}.
\label{eq:tr-hs}
\end{equation}
The Coulomb gas model can be converted following standard procedures
into a sine-Gordon model. The first step is to carry out a 
Hubbard-Stratonovich transformation on $\beta{\cal H}_{\rm C}$:
\begin{equation}
e^{-\beta{\cal H}_{\rm C}} = e^{\beta{\cal H}_{\rm L}}
\int{\cal D}\Phi e^{-\frac{1}{2}(\beta n^{2}K_{n})^{-1}
\int d\Omega \partial^{\alpha}\Phi
\partial_{\alpha}\Phi} e^{i\int d\Omega {\cal C}\Phi},
\end{equation}
where the Liouville factor $e^{\beta{\cal H}_{\rm L}}$ is needed to ensure
that $e^{-\beta{\cal H}_{\rm C}}$ be one when ${\cal C}=0$. 
Inserting this in Eq.~(\ref{eq:tr-hs}), we obtain
\begin{equation}
{\cal Z} = {\rm Tr}_{\rm v} y^{N}\int{\cal D}{\bf R}{\cal D}\Phi 
e^{-\beta{\cal H}_{\kappa}-\beta{\cal H}_{\Phi}}
e^{i\int d\Omega (s-\delta K)\Phi},
\end{equation}
where 
\begin{equation}
\beta{\cal H}_{\Phi} = \frac{1}{2}(\beta n^{2}K_{n})^{-1}\int d\Omega
\partial^{\alpha}\Phi\partial_{\alpha}\Phi.
\end{equation}
The only dependence on disclinations is now in the term linear in $n$. 
Thus to carry
out ${\rm Tr}_{\rm v}$, we need only to evaluate
\begin{eqnarray}
   &   & {\rm Tr}_{\rm v} y^{N}e^{i\int d\Omega s\Phi}   \nonumber   \\
   & = & \sum_{N_{+},N_{-}}\frac{1}{N_{+}!N_{-}!}\delta_{N_{+},N_{-}}
y^{N_{+}+N_{-}} \left( \int\frac{d\Omega}{a^{2}} 
e^{2\pi i\Phi(\Omega)/n}\right)^{N_{+}}  
\left( \int\frac{d\Omega}{a^{2}} 
e^{-2\pi i\Phi(\Omega)/n}\right)^{N_{-}}   \nonumber  \\
   & = & \sum_{N_{+},N_{-}}\frac{1}{N_{+}!N_{-}!}
\int \frac{d\omega}{2\pi} 
\left( y \int\frac{d\Omega}{a^{2}} 
e^{i \{ 2\pi [\Phi(\Omega)/n] -\omega \} }\right)^{N_{+}}    
\left( y \int\frac{d\Omega}{a^{2}} 
e^{-i \{ 2\pi [\Phi(\Omega)/n] -\omega \} }\right)^{N_{-}}     \nonumber  \\
   & = & \int \frac{d\omega}{2\pi} 
e^{(2y/a^{2})\int d\Omega \cos[2\pi(\Phi/n)-\omega]}.
\end{eqnarray}
Thus
\begin{equation}
{\cal Z} = \int \frac{d\omega}{2\pi}\int{\cal D}\Phi\int{\cal D}{\bf R}
               e^{-\beta{\cal H}_{\kappa}}e^{-\beta{\cal H}_{\Phi}}
               e^{(2y/a^{2})\int d\Omega
               \cos[2\pi(\Phi/n)-\omega]} e^{-i\int d\Omega
               \Phi \delta K}.
\label{eq:tr-sg}
\end{equation}
We can now change variables, letting $\Phi = (n/2\pi)(\Phi'+\omega)$.
The term linear in the Gaussian curvature then becomes
\begin{equation}
-i\int d\Omega \delta K \frac{n}{2\pi}(\omega + \Phi')
= -i\frac{p}{2\pi}\int d\Omega \Phi' \delta K,
\end{equation}
where we used $\int d\Omega \delta K = 0$.
The integral over $\omega$
in Eq.~(\ref{eq:tr-sg}) is now trivial, and dropping the prime we obtain
\begin{equation}
{\cal Z} = \int{\cal D}\Phi\int{\cal D}{\bf R} e^{-\beta{\cal H}_{\kappa}}
e^{-{\cal L}},
\label{eq:par-sg}
\end{equation}
where 
\begin{eqnarray}
{\cal L} & = & \frac{1}{2}(\beta n^{2}K_{n})^{-1}
\left(\frac{n}{2\pi}\right)^{2}
\int d\Omega \partial^{\alpha}\Phi\partial_{\alpha}\Phi 
\nonumber  \\  
        &   & -\frac{2y}{a^{2}}\int d\Omega \cos\Phi 
        -i\frac{n}{2\pi}\int d\Omega \Phi \delta K
\end{eqnarray}
is the sine-Gordon action on a fluctuating surface of genus zero.
The first two terms of this action are the gradient and cosine energies
present on a rigid sphere. The final term provides the principal coupling
between $\Phi$ and fluctuations in the metric.
It is analogous to the dilaton coupling \cite{Green-87} of string theory 
though here
the coupling constant is imaginary rather than real.
Note that the Liouville action is not explicitly present in 
Eq.~(\ref{eq:par-sg}). 
\par
In the regime $\beta\kappa \gg 1$, we can truncate the higher order terms
in $\rho$. In the normal gauge, the partition function becomes
\begin{eqnarray}
{\cal Z} & = & \int {\cal D}\rho{\cal D}\Phi
   \exp\left[-\frac{1}{2}\beta\kappa\int d\Omega
               \left((\nabla^{2}+2)\rho\right)^{2}  \right. \nonumber  \\
         &   & -\frac{1}{2}\beta\Gamma\int d\Omega (\nabla\Phi)^{2} +
   \frac{2y}{a^{2}}\int d\Omega\cos\Phi                   \nonumber  \\
         &   & \left. +i\frac{n}{2\pi}\int d\Omega \Phi(\nabla^{2}+2)\rho 
   \right]
\end{eqnarray}
where 
$\beta\Gamma \equiv 1/(4\pi^{2}\beta K_{n})$ and we used 
$\delta K = (\nabla^{2}+2)\rho$.      
To lowest order in $\rho$, the shape fluctuation field $\rho$ is linearly
coupled to the scalar field $\Phi$ which is the conjugate field to the
disclinations.
In Ref.~\cite{jp-lub}, we have shown the similar coupling in the
fluctuating flat membrane is quadratic in the shape fluctuation field.
\par
Integrating over the shape fluctuation field $\rho$ gives
the effective Hamiltonian for the conjugate field to the disclinations
\begin{eqnarray}
{\cal Z} & = & \int {\cal D}\Phi\exp\left[-\frac{1}{2}\beta\Gamma\int d\Omega
       \left((\nabla\Phi)^{2} + \mu^{2}\Phi^{2}\right) \right.  \nonumber  \\
         &   &  \left. + \frac{2y}{a^{2}}\int d\Omega\cos\Phi \right]
\end{eqnarray}
with $\mu^{2}=n^{2}K_{n}/\kappa$.
This is the massive sine-Gordon theory. The shape 
fluctuations induce the mass term for $\Phi$ field and screen the Coulombic
interaction between the disclinations giving the Yukawa interaction between
them.
This partition function is equivalent to that of the Yukawa gas Hamiltonian
on the rigid sphere with radius $R$;
\begin{eqnarray}
{\cal H}_{\rm Yukawa} & = & \frac{1}{2} \beta K_{A} \int d\Omega
s(\Omega) \frac{1}{(-\nabla^{2}+\mu^{2})} s(\Omega) \nonumber  \\
   & = & \frac{1}{2} \beta K_{A} \sum_{i,j} q_{i}q_{j}
G(\Omega_{i}-\Omega_{j}),
\end{eqnarray}
where
\begin{eqnarray}
G(\Omega_{i}-\Omega_{j}) & = & \sum_{l} 
\frac{2l+1}{l(l+1)+\mu^{2}}P_{l}(\cos\omega_{ij}) \nonumber  \\
   & = & -\frac{\pi}{\cos\left( (\nu+\frac{1}{2})\pi \right)}
P_{\nu}(-\cos\omega_{ij}),
\end{eqnarray}
where $P_{\nu}(\cdot)$ is the Legendre polynomial with degree 
$\nu=-1/2 \pm (\sqrt{1-4\mu^{2}})/2$ and
$\omega_{ij}$ is the angle between two disclinations at $\Omega_{i}$ and
$\Omega_{j}$.
For $0 \leq \mu \leq 1/2$, degree of Legendre polynomial, $\nu$, is real
and the length scale introduced by $\lambda_{d} \equiv (\mu/R)^{-1}$ is larger
than the system size, $2R$ after recovering the original length scale
by mapping the unit sphere back to the sphere with the radius $R$.
On the other hand, if $\mu > 1/2$, $\nu$ is a complex number,
$\nu=-1/2 \pm i\tau$ where $\tau = (\sqrt{4\mu^{2}-1})/2$ and
$\lambda_{d} < 2R$.
The length scale introduced by $\lambda_{d}=R/\mu$ may be interpreted 
as the Debye screening length arising from shape fluctuations.
\par
The interaction energy between two disclinations $i$ and $j$ at positions
$\Omega_{i}$ and $\Omega_{j}$ with strength $q_{i}$ and $q_{j}$ is given by
$q_{i}q_{j}G(d_{ij})$ where $d_{ij}=2R\sin(\omega_{ij}/2)$ is 
the chordal distance between two disclinations on the sphere with radius $R$.
The interaction $G(d_{ij})$ has the following limiting forms:
\begin{equation}
G(d_{ij}) \simeq \left\{ \begin{array}{ll}
        -\frac{1}{2} \ln \left( d_{ij}/2 \right), &
                  \mu^{-1} \gg 2,\; d_{ij} \ll 2R  \\
        -\frac{1}{2} \ln \left( d_{ij}/2 \right), &
                  \mu^{-1} \ll 2,\; d_{ij} \ll \lambda_{d}  \\
        e^{-d_{ij}/\lambda_{d}},  &
                  \mu^{-1} \ll 2,\; d_{ij} \gg \lambda_{d}
                         \end{array}
                 \right.
\end{equation}
Following the analogy of the 2D Coulomb gas, when the screening length is 
much larger than the system size, $\lambda_{d} \gg 2R$, the induced mass term 
arising from shape fluctuations is irrelavent for the KT transition
and the KT transition temperature is given by
$T_{\rm C} = \pi K_{n}/2$ for $\mu \ll 1/2$.
However, for $\mu \gg 1/2$, the screening length is 
shorter than the system size,
$\lambda_{d} \ll 2R$, and the mass term is relavent for the KT transition and
changes the universality class of the system.
There is no KT transition at non-zero temperature.
The disclinations are always unbound at non-zero temperature and
the KT transition temperature vanishes.
Thus, when $n^{2}K_{n}/\kappa \ll 1/4$, the effect of shape fluctuations
is irrelavent and the KT transition occurs at the finite temperature.
\section{Shape Changes Below $T_{\rm C}$}
For $n^{2}K_{n}/\kappa \ll 1/4$, there exists a KT transition at the finite 
temperature $T_{\rm C}=\pi K_{n}/2$. To describe the phase transition
and the corresponding shape changes, we introduce the magnitude of the
complex $n$-atic order parameter as in $\psi=\psi_{0} e^{in\Theta}$ and
consider the simplest long-wavelength Landau-Ginzburg Hamiltonian 
for the $n$-atic order parameter $\psi$;
\begin{equation}
{\cal H}_{\rm LG} = \int d^{2}u\sqrt{g} \left( r |\psi |^{2} +
\frac{1}{2}u |\psi |^{4} \right) + \frac{1}{2}K_{n}\int d^{2}u\sqrt{g}
g^{\alpha\beta}(\partial_{\alpha}-inA_{\alpha})\psi
(\partial_{\beta}+inA_{\beta})\psi^{*}.
\end{equation}
This Landau-Ginzburg $n$-atic Hamiltonian is similar to the Landau-Ginzburg
Hamiltonian for a superconductor in an external magnetic field,
\begin{equation}
{\cal H}_{\rm LG} = \int d^{3} x \left[ r | \psi |^{2} + C | ( {\bf \nabla} - 
ie^{*} {\bf A} )\psi|^2
+ \frac{1}{2}u | \psi |^{4} + \frac{1}{8\pi} ( {\bf \nabla} \times {\bf A}
- {\bf H} )^2 \right],
\end{equation}
where $e^{*}=2e/\hbar c$.
Both have a complex order
parameter $\psi$ with covariant derivatives providing a coupling between
$\psi$ and a ``vector potential" ${\bf A}$ or $A_{a}$.  
In a magnetic field, the
superconductor can undergo a second order mean-field transition from a normal
metal to the Abrikosov vortex lattice phase with a finite density of
vortices determined energetically by temperature and the magnetic field
${\bf H}$.  The magnetic field is conjugate to the vortex number $N_{\rm v}$
since $\int d^3 x ( {\bf \nabla} \times {\bf H} ) = L N_{\rm v} \phi_0$, 
where $L$ is the length of the sample along ${\bf H}$ and $\phi_0 = hc / 2e$ 
is the flux quantum. On a closed surface with $n$-atic order, 
there is a second-order mean-field transition to a
state with vortex number determined by topology rather than conjugate
external field.  Thus, the mean-field $n$-atic transition on a closed 
surface is analogous to the transition to an Abrikosov phase with a fixed 
number of vortices rather than fixed field conjugate to vortex number.
However, a Meissner phase with zero vortices does not exist because we do not 
have an analogue for magnetic intensity ${\bf H}$, rather topology fixes 
vortex number.
\par
The complete Hamiltonian for $n$-atic order on a deformable surface is 
${\cal H}={\cal H}_{\kappa}+{\cal H}_{\rm LG}$. 
The field $\rho(\Omega)$ in ${\bf R}(\Omega)=R_{0}(1+\rho(\Omega)){\bf e}_{r}$
can be expanded in spherical harmonics.
Any isotropic change in ${\bf R}$ can be described by $R_{0}$. In addition,
uniform translations, which change neither the shape nor the energy of the 
vesicle, correspond to distortions in $\rho$ with $l=1$ and can be absorbed 
by fixing the center of mass of the vesicle. These considerations imply that
$\rho$ will have no $l=0$ or $l=1$ components in spherical harmonic expansion:
\begin{equation}
\rho(\Omega)=\sum_{l=2}^{\infty}\sum_{m=-l}^{l}\rho_{lm}Y_{l}^{m}(\Omega).
\end{equation}
The shape and size of the vesicle are determined entirely by the parameters
$R_{0}$ and $\rho_{lm}$. The reduced tensors $\bar{g}_{\alpha\beta}$ and 
$\bar{K}_{\alpha\beta}$
do not depend on $R_{0}$. Therefore, $R_{0}$ can be expressed as a function
of ${\cal A}$ and $\rho_{lm}$ via the relation 
\begin{eqnarray}
{\cal A} & = & \int d^{2}u\sqrt{g}  = R_{0}^{2}\int d^{2}u\sqrt{\bar{g}}   
\nonumber \\
  & = & R_{0}^{2}\int d\Omega \left[ (1+\rho)^{2} + \frac{1}{2}\left( 
        (\partial_{\theta}\rho)^{2} + \frac{(\partial_{\phi}\rho)^{2}}
        {\sin^{2}\theta} \right) + O(\rho^{4}) \right]     \nonumber  \\
  & = & 4\pi R_{0}^{2} \left[ 1 + \frac{1}{2\cdot 4\pi}\sum_{l=2}^{\infty}
        \sum_{m}(l(l+1)+2)|\rho_{lm}|^{2} + O(\rho^{4})  \right] .
\end{eqnarray}
In the disordered phase above $T_{\rm C}$, $\rho=0$ and 
$R=R_{0}$. We will use the Hamiltonian ${\cal H}$ expressed in terms of
reduced parameters and the constant area ${\cal A}$ in our calculations of
shape changes below the second-order disordered-to-$n$-atic transition.
\par
Using the spherical polar coordinates with origin at the center of mass of 
the vesicle, the metric tensor $g_{\alpha\beta}$ can be written as
\begin{eqnarray}
g_{\alpha\beta} & = & \partial_{\alpha}{\bf R}\cdot\partial_{\beta}{\bf R} = 
                         R_{0}^{2}\bar{g}_{\alpha\beta}   \\
\bar{g}_{\alpha\beta} & = & (1+\rho)^{2}\bar{g}^{0}_{\alpha\beta}+
                         (\partial_{\alpha}\rho)(\partial_{\beta}\rho)   \\
\bar{g}^{0}_{\alpha\beta} & = & \left( \begin{array}{cc}
                         1 & 0 \\
                         0 & \sin^{2}\theta
                         \end{array} \right)
\end{eqnarray}
and the curvature tensor $K_{\alpha\beta}$ is
\begin{eqnarray}
K_{\alpha\beta} & = & {\bf N}\cdot D_{\alpha}D_{\beta}{\bf R} = 
R_{0}\bar{K}_{\alpha\beta}  \\
\bar{K}_{\alpha\beta} & = & \frac{1}{\sqrt{1+(\nabla\rho)^{2}}}
                           \left(  \begin{array}{cc}
   (1+\rho)-\partial_{\theta}^{2}\rho + 2(\partial_{\theta}\rho)^{2} & 
   -\sin\theta\partial_{\theta}\frac{\partial_{\phi}\rho}{\sin\theta} +
            2\partial_{\theta}\rho\partial_{\phi}\rho     \\
   -\sin\theta\partial_{\theta}\frac{\partial_{\phi}\rho}{\sin\theta} +
            2\partial_{\theta}\rho\partial_{\phi}\rho  &   
   (1+\rho)\sin^{2}\theta-\sin\theta\cos\theta\partial_{\theta}\rho
            -\partial_{\phi}^{2}\rho+2(\partial_{\phi}\rho)^{2} 
                           \end{array}  \right),
\end{eqnarray}
where $(\nabla\rho)^{2} = (\partial_{\theta}\rho)^{2} + 
\frac{1}{\sin^{2}\theta}(\partial_{\phi}\rho)^{2}$.
The Hamiltonian is a functional of $\psi(\Omega)$ and $\rho(\Omega)$
at the fixed area ${\cal A}$. To find the equilibrium form of $\psi(\Omega)$
and $\rho(\Omega)$, we need to minimize ${\cal H}$ over $\psi(\Omega)$ and 
$\rho(\Omega)$. There is a considerable simplification if we restrict our
attention to the neighborhood of the transition temperature $T_{\rm C}$.
Near $T_{\rm C}$, to order $\psi^{4}$, the variation of the curvature energy
is of order $\kappa\rho^{2}$. When this is comparable to the Landau-Ginzburg
free energy for $\psi$, we only need to keep couplings of order $\rho\psi^{2}$.
To this order, the Landau-Ginzburg Hamiltonian for $\psi$ becomes
\begin{eqnarray}
{\cal H}_{\rm LG} & = & \int d\Omega \left( r |\psi |^{2} +
    \frac{1}{2}u |\psi |^{4} \right) + \frac{1}{2}K_{n}\int d\Omega
    g^{\alpha\beta}(\partial_{\alpha}-inA_{\alpha}^{0})\psi
    (\partial_{\beta}+inA_{\beta}^{0})\psi^{*}  \nonumber  \\
  & + & \int d\Omega 2\rho \left( r |\psi |^{2} + \frac{1}{2}u
    |\psi |^{4} \right) + \int d\Omega J^{\alpha}\delta A_{\alpha},
\end{eqnarray}
where $A_{\theta}^{0}=0$ and $A_{\phi}^{0}=-\cos\theta$.
In this formula, 
\begin{eqnarray}
J^{\alpha} & = & g^{-1/2}\frac{\delta{\cal H}}{\delta A_{\alpha}} \nonumber \\
      & = & \frac{nK_{n}}{2i}\bar{g}^{0 \alpha\beta}
            (\psi\partial_{\beta}\psi^{*}
            -\psi^{*}\partial_{\beta}\psi + 2in A^{0}_{\beta} |\psi |^{2})
\end{eqnarray}
has been evaluated at $\rho_{lm}=0$, $R_{0}=R$
and $\delta A_{\alpha} = \gamma_{\alpha}^{\beta}\partial_{\beta}\rho$.
In terms of $\rho$, the curvature energy can be written as
\begin{equation}
{\cal H}_{\kappa}=\frac{1}{2}\kappa\int d\Omega \left( (\nabla^{2}+2)\rho
\right)^{2},
\end{equation}
where 
\begin{equation}
\nabla^{2}\rho=-\left( \frac{1}{\sin\theta}
\partial_{\theta}\sin\theta\partial_{\theta}\rho
+\frac{1}{\sin^{2}\theta}\partial_{\phi}^{2}\rho \right).
\end{equation}
Hence, to order $\psi^{4}$, we can write ${\cal H}$ as ${\cal H}={\cal H}^{0}
+\delta {\cal H}$ with
\begin{eqnarray}
{\cal H}^{0} & = & \frac{{\cal A}}{4\pi}\int d\Omega \left( r |\psi |^{2} +
       \frac{1}{2}u |\psi |^{4} \right)   \nonumber  \\
  & + & \frac{1}{2}K_{n}\int d\Omega \bar{g}^{0\alpha\beta}
       (\partial_{\alpha}-inA_{\alpha}^{0})
       \psi (\partial_{\beta}+inA_{\beta}^{0})\psi^{*},
\end{eqnarray}
which does not depend on $\rho(\Omega)$ and
\begin{eqnarray}
\delta {\cal H} & = & \frac{{\cal A}}{4\pi}\int d\Omega 2\rho 
             \left( r |\psi |^{2} +  \frac{1}{2}u |\psi |^{4} \right) 
\nonumber \\
  & - & \int d\Omega \frac{\rho}{\sin\theta} \partial_{\alpha}
        (\gamma^{\alpha}_{\beta} J^{\beta})   \nonumber  \\
  & + & \frac{1}{2}\kappa\int d\Omega \left( (\nabla^{2}+2)\rho \right)^{2},
\end{eqnarray}
which contains $\rho(\Omega)$.
This can be rewritten as
\begin{equation}
{\cal H}= {\cal H}^{0} + \int d\Omega\rho(\Omega) \tilde{\Phi}(\Omega) + 
\frac{1}{2}\kappa\int d\Omega d\Omega' \rho(\Omega) M(\Omega,\Omega')
\rho(\Omega')
\end{equation}
with
\begin{equation}
\tilde{\Phi}(\Omega) = \frac{{\cal A}}{2\pi}
             \left( r |\psi |^{2} +  \frac{1}{2}u |\psi |^{4} \right)
- \frac{1}{\sin\theta} \partial_{\alpha}(\gamma^{\alpha}_{\beta} J^{\beta})
\end{equation}
\begin{equation}
M(\Omega,\Omega') = \sum_{l=2}^{\infty}\sum_{m=-l}^{l}
Y_{l}^{m*}(\Omega) (l-1)^{2}(l+2)^{2} Y_{l}^{m}(\Omega').
\end{equation}
Minimizing over $\rho(\Omega)$, we obtain
\begin{equation}
\rho(\Omega)=-\frac{1}{\kappa}\int d\Omega' M^{-1}(\Omega,\Omega')
\tilde{\Phi}(\Omega'),
\end{equation}
where $M^{-1}(\Omega,\Omega')$ satisfies
\begin{equation}
\int d\Omega_{3} M(\Omega_{1},\Omega_{3}) M^{-1}(\Omega_{3},\Omega_{2})
= \delta_{(l\geq 2)}(\Omega_{1}-\Omega_{2})
\end{equation}
and $\delta_{(l\geq 2)}(\Omega_{1}-\Omega_{2})$ is the Dirac delta function in
spherical harmonics space with $l\geq 2$.
By substituting this for $\rho(\Omega)$ into ${\cal H}$, we have
\begin{eqnarray}
{\cal H}_{\rm eff} & = & \frac{{\cal A}}{4\pi}\int d\Omega 
       \left( r |\psi |^{2} + \frac{1}{2}u |\psi |^{4} \right)   \nonumber  \\
  & + & \frac{1}{2}K_{n}\int d\Omega \bar{g}^{0\alpha\beta}
       (\partial_{\alpha}-inA_{\alpha}^{0})
       \psi (\partial_{\beta}+inA_{\beta}^{0})\psi^{*}   \nonumber  \\
  & - & \frac{1}{2\kappa}\sum_{l=2}^{\infty}\sum_{m=-l}^{l}
       \frac{|\tilde{\Phi}_{lm}|^{2}}{(l-1)^{2}(l+2)^{2}},
\end{eqnarray}
where
\begin{eqnarray}
\tilde{\Phi}_{lm} & = & \int d\Omega Y_{l}^{m*}(\Omega)\tilde{\Phi}(\Omega) \\
\tilde{\Phi}(\Omega) & = & \frac{{\cal A}r_{c}}{2\pi} |\psi |^{2} -
  \frac{1}{\sin\theta} \partial_{\alpha}(\gamma^{\alpha}_{\beta} J^{\beta}).
\end{eqnarray}
\par
Before proceeding with our analysis of the $n$-atic transition on a
deformable sphere, it is useful to recall Abrikosov's calculation of the
transition to the vortex state.  The first step is to calculate the
eigenfunctions of the harmonic part of ${\cal H}_{\rm GL}$ when 
${\bf \nabla} \times {\bf A}= {\bf H}$.  
These can be divided into highly degenerate sets separated by
an energy gap, $\hbar \omega_c = 2C e^{*} H$.  In the Landau gauge with 
${\bf A} = (0,Hx,0)$, the eigenfunctions in the lowest energy manifold are 
$\psi_k = e^{iky} e^{- e^{*} H( x - x_k )^2 }$ where $x_k = k/ e^{*} H$.  
The order parameter $\psi ({\bf x})$ of the ordered state is expressed as 
a linear combination $\psi ({\bf x}) = \sum C_k \psi_k$, where the complex
parameters $C_k$ are determined by minimization of ${\cal H}_{\rm GL}$.
\par
With this analogy in mind, we will follow Abrikosov's treatment of the 
superconducting transition near $H_{{\rm C}_{2}}$ to study the development of
$n$-atic order on a sphere.
We first diagonalize ${\cal H}$ in the harmonic level when $\rho=0$ and
$A_{\theta}=A_{\theta}^{0}$, $A_{\phi}=A_{\phi}^{0}$, that is,
we determine the functions $\psi$ that satisfy $K_{A}D_{\alpha}D^{\alpha}\psi=
\varepsilon\psi$, for $\rho=0$, $A_{\alpha}=A_{\alpha}^{0}$. 
Nonlinearities arise from the $\mid\psi\mid^{4}$ term in Landau-Ginzburg
Hamiltonian for $\psi$ and from the fact that $A_{\alpha}$ and 
$g_{\alpha\beta}$ depend
nonlinearly on the deviation from the ideal spherical shape.
To minimize ${\cal H}_{\rm eff}$, 
first we seek the lowest energy configurations
of the operators corresponding to the harmonic terms only. Then, we 
linearly combine these eigenfunctions to get the function which has the lowest
energy for ${\cal H}_{\rm eff}$. That is, the manifold of the lowest energy
eigenstates of the harmonic terms is highly degenerate. Nonlinear terms 
pick out the lowest energy state which is the linear combination of the
degenerate eigenstates of the harmonic terms and resolve the degeneracy.
We divide ${\cal H}_{\rm eff}$ into the harmonic part, ${\cal H}_{\rm har}$,
and the nonlinear part, ${\cal H}_{\rm nl}$:
\begin{eqnarray}
{\cal H}_{\rm har} & = & \frac{{\cal A}}{4\pi}\int d\Omega r |\psi |^{2}
       + \frac{1}{2}K_{n}\int d\Omega \bar{g}^{0\alpha\beta}
       (\partial_{\alpha}-inA_{\alpha}^{0})
       \psi (\partial_{\beta}+inA_{\beta}^{0})\psi^{*}   \\
{\cal H}_{\rm nl} & = &  \frac{{\cal A}}{4\pi}\int d\Omega \frac{1}{2}u
                         |\psi |^{4}
       - \frac{1}{2\kappa}\sum_{l=2}^{\infty}\sum_{m=-l}^{l}
       \frac{|\tilde{\Phi}_{lm}|^{2}}{(l-1)^{2}(l+2)^{2}}
\end{eqnarray}
The differential operator for ${\cal H}_{\rm har}$ is
\begin{equation}
\frac{1}{\sqrt{\bar{g}^{0}}}\frac{\delta {\cal H}_{\rm har}}{\delta\psi^{*}} 
      \equiv  (\frac{{\cal A}}{4\pi} + \frac{K_{n}}{2}\triangle ) \psi ,
\end{equation}
where
\begin{equation}
\triangle = - \frac{1}{\sqrt{\bar{g}^{0}}}
  (\partial_{\alpha}-inA_{\alpha}^{0})\sqrt{\bar{g}^{0}}\bar{g}^{0\alpha\beta}
  (\partial_{\beta}-inA_{\beta}^{0}) .
\end{equation}
In the spherical polar representation, $\triangle$ has the form
\begin{equation}
\triangle = - \frac{1}{\sin\theta}\left[ \partial_{\theta}
    (\sin\theta \partial_{\theta}) + \frac{1}{\sin\theta}\partial_{\phi}^{2}
    - n^{2}\frac{\cos^{2}\theta}{\sin\theta} + 2in\frac{\cos\theta}{\sin\theta}
    \partial_{\phi} \right].
\end{equation}
One possible form of the eigenfunctions of $\triangle$ has 
the structure of $\sin^{n}\theta$.
\begin{equation}
\triangle \sin^{n}\theta = n \sin^{n}\theta.
\end{equation}
The complete spectrum and the eigenfunctions are derived in the Appendix.
Introducing the projection representation, we can reparametrize the sphere
using a $(z,\bar{z})$ coordinate system defined by
\begin{equation}
z = \tan\frac{\theta}{2}e^{i\phi} \; ; \;\; 
\bar{z} = \tan\frac{\theta}{2}e^{-i\phi}.
\end{equation}
The inverse transformation gives
\begin{equation}
\theta = 2\tan^{-1}\sqrt{z\bar{z}} \; ; \;\; 
\phi = \frac{1}{2i}\ln\frac{z}{\bar{z}}.
\end{equation}
The corresponding metric tensor, spin connection, and the differential 
operator, $\triangle$, in this 
representation can be written as follows
\begin{equation}
\bar{g}^{0ab} = \frac{2}{(1+|z|^{2})^{2}} \left( \begin{array}{cc}
                                                   0 & 1 \\
                                                   1 & 0
                                                   \end{array} \right)
\end{equation}
\begin{equation}
\Omega_{z}^{0} = -\frac{1}{2iz}\frac{1-|z|^{2}}{1+|z|^{2}}  \;\; ; \;\;\;
\Omega_{\bar{z}}^{0} = \frac{1}{2i\bar{z}}\frac{1-|z|^{2}}{1+|z|^{2}}
\end{equation}
\begin{equation}
\triangle = - (1+|z|^{2})^{2} [ (\partial_{z}\partial_{\bar{z}}
    - A_{z}^{0} A_{\bar{z}}^{0}) - 2in (A_{z}^{0}
    \partial_{\bar{z}} + A_{\bar{z}}^{0}\partial_{z}) ].
\end{equation}
The eigenfunction $\sin^{n}\theta$ can be written as
\begin{equation}
f(|z|^{2}) = \left( \frac{2|z|}{1+|z|^{2}} \right)^{n}.
\end{equation}
This eigenfunction has vortices at $|z| = 0$ and $|z| = \infty$. Now,
the vortices can be arbitrarily placed by multiplying $Q(z)$ which is
a function of $z$ alone. From the differential operator $\triangle$, we see
\begin{eqnarray}
\triangle f(|z|^{2}) & = & n f(|z|^{2})  \nonumber  \\
\Rightarrow \triangle f(|z|^{2})Q(z) & = & n f(|z|^{2})Q(z).
\end{eqnarray}
Thus, the full eigenfunctions of $\triangle$ are the product of
$f(|z|^{2})$ and the arbitrary function $Q(z)$.
In particular, for $n=1$, we have
\begin{equation}
f(|z|^{2}) = \frac{2|z|}{1+|z|^{2}}.
\end{equation}
By taking $Q(z) = \frac{(z-z_{1})(z-z_{2})}{z}$, 
\begin{equation}
\psi = ({\rm Const.})\frac{2|z|}{1+|z|^{2}}\frac{(z-z_{1})(z-z_{2})}{z} ,
\end{equation}
which means that $\psi$ has the vortices at $z=z_{1}$ and $z=z_{2}$ and 
that $\psi$ goes to $z_{1}z_{2}e^{-i\phi}$ as $z$ goes 0, $\psi$ goes to 
$e^{i\phi}$ as $z$ goes to $\infty$.
This means there is an enormous degeneracy in the lowest energy
manifold of the harmonic terms.
In the lowest energy manifold of the harmonic terms, $\psi$ will have
exactly $2n$ zeros specifying the vortex positions, each of strength $1/n$.
\par
In view of this vortex strength and distribution,
we choose $Q(z)$ as
\begin{equation}
Q(z) = \frac{\alpha}{z^{n}}\prod_{i=1}^{2n}(z-z_{i}),
\end{equation}
where $\alpha,z_{i}$ are arbitrary and $z_{i} \neq z_{j}$ if $i \neq j$.
By minimizing ${\cal H}_{\rm eff}$ with the degenerate eigenfunctions of 
$\triangle$,
\begin{equation}
\psi = \alpha \left( \frac{2|z|}{1+|z|^{2}} \right)^{n} \frac{1}{z^{n}}
       \prod_{i=1}^{2n}(z-z_{i}),
\end{equation}
the non-linear terms pick out the set 
$\{z_{i} \mid 1 \leq i \leq 2n \}$ that gives the lowest energy for 
${\cal H}_{\rm eff}$.
This set $\{z_{i}\}$ also gives $|\psi|$, vortex positions, and the 
shape change ${\bf R}$ of the membrane. In terms of spherical coordinates,
\begin{equation}
\psi = ({\rm Const.}) \prod_{i=1}^{2n} \left(
       \sin\frac{\theta}{2}\cos\frac{\theta_{i}}{2}e^{i(\phi-\phi_{i})/2} -
       \cos\frac{\theta}{2}\sin\frac{\theta_{i}}{2}e^{-i(\phi-\phi_{i})/2}
       \right) \equiv \psi_{0} P(\Omega)
\label{psi-eigen}
\end{equation}
and
\begin{equation}
|\psi|^{2} = \frac{\psi_{0}^{2}}{{\cal N}} \prod_{i=1}^{2n}(1 - \cos\gamma_{i})
\end{equation}
where $\cos\gamma_{i}=\cos\theta\cos\theta_{i}+
       \sin\theta\sin\theta_{i}\cos(\phi-\phi_{i})$, $\Omega_{i}=(\theta_{i},
\phi_{i})$ specifies the position of the $i$-th zero of $\psi$, and
${\cal N}$ is the normalization factor which makes
\begin{equation}
\int d\Omega |\psi|^{2} = \psi_{0}^{2}.
\end{equation}
Since $\psi$ of Eq.~(\ref{psi-eigen})
is the most general function in the lowest energy
manifold, the order parameter and vesicle shape just below the transition
temperature are obtained by minimizing ${\cal H}_{\rm eff}$ over $\psi_{0}$ 
and the positions of zeros.
Inserting this expression for $\psi$, the free energy becomes 
\begin{eqnarray}
{\cal H}_{\rm eff} & = & \frac{{\cal A}}{4\pi}\int d\Omega 
       \left( (r-r_{c}) |\psi |^{2} + \frac{1}{2}u |\psi |^{4} \right)   
\nonumber  \\
  & - & \frac{1}{2\kappa}\sum_{l=2}^{\infty}\sum_{m=-l}^{l}
       \frac{|\tilde{\Phi}_{lm}|^{2}}{(l-1)^{2}(l+2)^{2}} ,
\end{eqnarray}
where $r_{c}=-4\pi nK_{A}/{\cal A}$ and
\begin{eqnarray}
\tilde{\Phi}(\Omega) & = & -nK_{n}\psi_{0}^{2}\prod_{i=1}^{2n}
   (1-\cos\gamma_{i}) \left[ 1 + \sum_{i=1}^{2n}\frac{\cos\gamma_{i}}
   {(1-\cos\gamma_{i})} +  
   \sum_{i>j}\frac{\cos\gamma_{ij}-\cos\gamma_{i}\cos\gamma_{j}}
   {(1-\cos\gamma_{i})(1-\cos\gamma_{j})}  \right]   \nonumber  \\
                     & \equiv &  \psi_{0}^{2}\Phi(\Omega).
\end{eqnarray}
Minimization over $\psi_{0}$ leads to 
$\psi_{0}^{2} = -r[\{\Omega_{i}\}]/u[\{\Omega_{i}\}]$ 
and the effective free energy density
\begin{equation}
f[\{\Omega_{i}\}] = -\frac{1}{2}\frac{r^{2}[\{\Omega_{i}\}]}
    {u[\{\Omega_{i}\}]},
\end{equation}
depending only on the positions $\{\Omega_{i}\}$ of the zeros of $\psi(\omega)$
where
\begin{eqnarray}
r[\{\Omega_{i}\}] & = & (r-r_{c})\int d\Omega | P(\Omega) |^{2}  \\
u[\{\Omega_{i}\}] & = & u\int d\Omega | P(\Omega) |^{4} - \frac{4\pi}
     {\kappa{\cal A}}\sum_{l=2}^{\infty}\sum_{m=-l}^{l}
       \frac{|\tilde{\Phi}_{lm}|^{2}}{(l-1)^{2}(l+2)^{2}}.
\end{eqnarray}
\par
\begin{figure}
\centerline{\psfig{figure=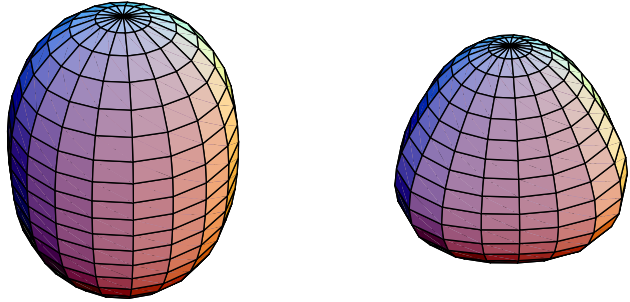}}
\caption{Mean-field shapes of deformable surfaces of genus zero with
vector order ($n=1$) on the left and nematic order ($n=2$) on the right. 
Above the mean-field transition temperature, the equilibrium
shape is spherical for all $n$. Below the transition temperature, the
equilibrium shape depends on $n$ and has a  ellipsoidal form
with $2$ ground state vortices with strength 1 located at opposite poles 
for $n=1$ and a tetrahedral form with $4$ ground state vortices with 
strength $1/2$ located at the vertices of a tetrahedron for $n=2$.
}
\label{fig12}
\end{figure}
\begin{figure}
\centerline{\psfig{figure=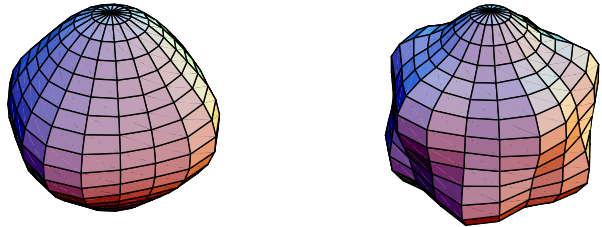}}
\caption{Mean-field shapes of deformable surfaces of genus zero with
tertic order ($n=3$) on the left and hexatic order ($n=6$) on the right. 
Below the transition temperature, the equilibrium shape has a octahedral form
with $6$ ground state vortices with strength $1/3$
located at the vertices of a octahedron for $n=3$ and
a icosahedral form with $12$ ground state vortices with strength $1/6$
located at the vertices of a icosahedron for $n=6$.
}
\label{fig36}
\end{figure}
\begin{figure}
\centerline{\psfig{figure=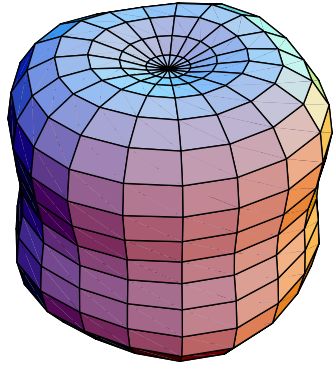}}
\caption{Mean-field shapes of deformable surfaces of genus zero with
quartic order ($n=4$). Below the transition temperature, the
equilibrium shape has a distorted cubic form
with $8$ ground state vortices with strength $1/4$ located at 
the vetices of a distorted cube obtained by rotating its top face
about its 4-fold axis by $\pi/4$ and compressing opposite faces.
}
\label{fig4}
\end{figure}

The next step is to minimize $f[\{\Omega_{i}\}]$ over $\Omega_{i}$. Shape 
changes are described by $\rho(\Omega)=\sum_{l=2}^{\infty}\sum_{m=-l}^{l}
\rho_{lm}Y_{m}^{l}(\Omega)$ where
\begin{eqnarray}
\rho_{lm} & = & -\frac{1}{\kappa (l-1)^{2}(l+2)^{2}}
                     \int d\Omega Y_{l}^{m*}(\Omega)\tilde{\Phi}(\Omega)  
\nonumber \\
          & \equiv & -\frac{\psi_{0}^{2}}{\kappa (l-1)^{2}(l+2)^{2}}
                     \Phi_{lm}(\{\Omega_{i}\}) ,
\end{eqnarray}
with $\{\Omega_{i}\}$ that minimize $f[\{\Omega_{i}\}]$.
In the disordered phase above the transition temperature, $\psi_{0}=0$.
Hence, $\rho=0$ and $R_{0}=R$. The vesicle shape is spherical. In the ordered
phase below the transition temperature, we have been able to evaluate
$f[\{\Omega_{i}\}]$ analytically for $n=1$ and $n=2$. For these two cases,
we find as expected that the minimum energy configurations are, respectively,
those with zeros of $\psi(\Omega)$ at opposite poles and at the vertices of
a tetrahedron. The shape function $\rho(\Omega)=(nK_{n}/\kappa)\psi_{0}^{2}
\bar{\rho}^{(n)}(\Omega)$ associated with $n$-atic order is 
proportional to $\psi_{0}^{2} \sim (r-r_{c})$ to the order of our 
calculations. In general, the Legendre 
decomposition of $\bar{\rho}^{(n)}$ will contain Legendre polynomials of
order $2n$. For $n=1$ and $n=2$, we find
\begin{eqnarray}
\bar{\rho}^{(1)}(\Omega) & = & \frac{1}{6}\sqrt{\frac{\pi}{5}} 
                       Y_{2}^{0}(\Omega),     
\label{shape1} \\
\bar{\rho}^{(2)}(\Omega) & = & \frac{8}{135}\sqrt{\frac{\pi}{7}}
        \left[ Y_{3}^{0}(\Omega) - \sqrt{\frac{2}{5}}(Y_{3}^{3}(\Omega)
        -Y_{3}^{-3}(\Omega)) \right]  \nonumber  \\
                         &  & + \frac{4}{3645}\sqrt{\pi}
        \left[ Y_{4}^{0}(\Omega) + \sqrt{\frac{10}{7}}(Y_{4}^{3}(\Omega)
        -Y_{4}^{-3}(\Omega)) \right].
\label{shape2}
\end{eqnarray}
Figure~\ref{fig12} shows the shapes described by these functions. 
The transformations
from the initial spherical shape to distorted shapes occur continuously.
Our calculations for the shape are valid to order $(r-r_{c})$. As
temperature is lowered, higher-order terms in $(r-r_{c})$ and higher-order 
spherical harmonics are needed to describe the equilibrium shape. 
In Ref. \cite{MacKintosh-91}, 
a variational function for $\psi$ and spherical harmonics up
to order $8$ were used to calculate the shape for $n=1$ for temperatures
well below the transition. 
We use symmetry considerations to treat $n=3,4,$ and $6$. For $n=3$ and $n=6$,
we expect the zeros of $\psi(\Omega)$ to lie, respectively, at the vertices of
an octahedron and an icosahedron. We find the shape functions for $n=3$ and 
$n=6$
\begin{eqnarray}
\bar{\rho}^{(3)}(\Omega) & = & \frac{1}{297}\sqrt{\pi}
        \left[ Y_{4}^{0}(\Omega) + \sqrt{\frac{5}{14}}(Y_{4}^{4}(\Omega)
        +Y_{4}^{-4}(\Omega)) \right]   \nonumber  \\
                         &  & + \frac{1}{60060}\sqrt{13\pi}
        \left[ Y_{6}^{0}(\Omega) - \sqrt{\frac{7}{2}}(Y_{6}^{4}(\Omega)
        +Y_{6}^{-4}(\Omega)) \right],    \\
\bar{\rho}^{(6)}(\Omega) & = & \frac{15488}{275559375}
\sqrt{13\pi} \left[ Y_{6}^{0}(\Omega) + \sqrt{\frac{7}{11}} 
( Y_{6}^{5}(\Omega) - Y_{6}^{-5}(\Omega) ) \right]       \nonumber \\
   &  & + \frac{256}{692803125}\sqrt{21\pi}\left[
Y_{10}^{0}(\Omega) + \sqrt{\frac{33}{13}} 
( Y_{10}^{5}(\Omega) - Y_{10}^{-5}(\Omega) ) 
+ \frac{187}{247}( Y_{10}^{10}(\Omega) + Y_{10}^{-10}(\Omega) ) 
\right]  \nonumber   \\
   &  & - \frac{512}{6834953125} \sqrt{\pi}\left[
Y_{12}^{0}(\Omega) - \frac{1}{3}\sqrt{\frac{286}{119}} 
( Y_{12}^{5}(\Omega) - Y_{12}^{-5}(\Omega) ) 
+ \frac{247}{357}( Y_{12}^{10}(\Omega) + Y_{12}^{-10}(\Omega) ) 
\right] 
\end{eqnarray}
Figure~\ref{fig36} shows the shapes described by these functions. For $n=4$,
following the calculations of Palffy-Muhoray \cite{Palffy}, 
we expect the zeros of
$\psi(\Omega)$ to lie at the vertices of a distorted cube obtained by 
rotating its top face about its fourfold axis by $\pi/4$ and compressing
opposite faces. For this case, we minimized the energy over a single
parameter determining the separation between opposite rotated faces and obtain
\begin{eqnarray}
\bar{\rho}^{(4)}(\Omega)  & =  & -0.0184631 Y_{2}^{0}(\Omega)
         -0.305216 Y_{4}^{0}(\Omega)
         0.0335485 (Y_{5}^{4}(\Omega)+Y_{5}^{-4}(\Omega))  \nonumber     \\
                       &  & + 0.0201099 Y_{6}^{0}(\Omega)
         + 0.00721177 (Y_{7}^{4}(\Omega)+Y_{7}^{-4}(\Omega))  \nonumber   \\
                       &  &  - 0.00159671 \left[ Y_{8}^{0} - 0.578489
          (Y_{8}^{8}(\Omega)+Y_{8}^{-8}(\Omega))
         \right].  
\end{eqnarray}
Figure~\ref{fig4} shows the shape for $n=4$.
\section{Conclusions}
We have presented the general $n$-atic Hamiltonian in terms of the $n$-th
rank spherical tensors and have shown there exists an effective KT transition
 for $n^{2}K_{n}/\kappa \ll 1/4$.
Also we have presented here an analysis of the mean-field transition 
to $n$-atic order on a fixed-area surface of genus zero and the corresponding
continuous shape changes below the KT transition temperature in this 
parameter region.  
We are really dealing with two kinds of
order: $n$-atic order and the positional order of vortices.  In mean-field
theory, these two kinds of order develop simultaneously.  
Our analysis is very similar to that of
Abrikosov for the transition from a normal metal to a vortex lattice in a
type II superconductor at $H_{{\rm C}_{2}}$ 
and ignores the effects of fluctuations.
In superconductors, fluctuations drive the normal-to-superconducting transition
first order in a nonvanishing magnetic field \cite{Fetter-69}.  
In the Abrikosov phase, fluctuations of the vortex
lattice destroy superconductivity but not long-range periodic order.  In
two-dimensions, screening of vortices drives the Kosterlitz-Thouless
transition in an infinite superconductor in zero field to zero.  
Both of the above effects may be important for $n$-atic order on a
sphere. 
We have been able to find in the parameter region $n^{2}K_{n}/\kappa \ll
1/4$, the effect of shape fluctuations is irrelavent and will not
lead to qualitative changes in our results.
However, outside this region, {\it i.e.} for $n^{2}K_{n}/\kappa \gg 1/4$,
shape fluctuations are relavent. 
The interaction between the vortices is screened
to become the massive sine-Gordon theory and hence is equivalent to
a neutral Yukawa gas on a sphere. Consequently, this screening effect
derives the Kosterlitz-Thouless transition temperature to zero. The vortices 
are always unbound for a non-zero temperature. 
\par
In Ref. \cite{Evans-96}, Evans has discussed the shape changes of the
$n$-atic Hamiltonian in the opposite limit (high temperature limit) to the
mean-field limit described in this paper. Using the lowest Landau level
approximation, the effect of thermal fluctuations are discussed.
Although we agree with him in that the amplitudes of deformations have
zero thermal average, since the lowest Landau levels include only the 
ground state vortices we are not sure if this approximation is valid in the 
high temperature phase where thermally excited vortex-antivortex pairs are
always unbound. 
Thus we believe the effects of fluctuations to the shape changes in the 
disordered state deserve further investigation.

\appendix
\section{Complete Spectrum of $\Delta$}

We present the complete spectrum of the differential operator, $\triangle$,
for the harmonic part, ${\cal H}_{\rm har}$. The operator $\triangle$
in the spherical polar representation,
\begin{equation}
\triangle = -\frac{1}{\sin\theta}\left( 
             \partial_{\theta} \sin\theta \partial_{\theta} +
             \frac{1}{\sin\theta}(\partial_{\phi} + in\cos\theta)^{2} \right),
\end{equation}
is similar to the differential operator $-({\bf \nabla}-i{\bf A})^{2}$
for the superconductivity where
\begin{equation}
{\bf \nabla} =  {\bf \theta}\partial_{\theta} + 
                   {\bf \phi}\frac{1}{\sin\theta}\partial_{\phi} ,
\end{equation}
and
\begin{equation}
{\bf A} =  {\bf \theta} A_{\theta}+ {\bf \phi} A_{\phi}.
\end{equation}
The Schr\"{o}dinger equation on a sphere in the field of a Dirac magnetic 
monopole of strength $g$ at the origin plus an infinitely long and thin 
solenoid carrying flux $F$ along the $z$ axis is described by
\begin{equation}
\frac{1}{2}({\bf \nabla}-i{\bf A})^{2}\psi = E\psi
\end{equation}
with
\begin{equation}
A_{\theta}=0 \; ; \;\; A_{\phi} = \frac{g}{\sin\theta}(1-\cos\theta) +
\frac{F}{2\pi\sin\theta}.
\end{equation}
The monopole has been chosen to have a Dirac string singularity along
the negative $z$ axis. Setting $g=n$, $F=-2\pi n$, we find
\begin{equation}
A_{\theta}=0 \; ; \;\; A_{\phi}=-\frac{n\cos\theta}{\sin\theta}
\end{equation}
and $-({\bf \nabla}-i{\bf A})^{2}$ is exactly identical to $\triangle$.
Hence, the differential operator, $\triangle$, is equivalent to the 
Hamiltonian describing the field of a Dirac magnetic monopole of strength
$n$ at the origin plus an infinitely long and thin solenoid carrying flux 
$2\pi n$ along the positive $z$ axis.
Following the method in Ref. \cite{Roy-83}, 
we define the formal differential operator
\begin{equation}
{\bf J} \equiv -i{\bf r} \times ({\bf \nabla}-i{\bf A}) -n{\bf r}
\end{equation}
with components
\begin{eqnarray}
J_{\pm} & = & J_{x} \pm i J_{y} = e^{\pm i\phi} \left[
              \pm\partial_{\theta}+i\cot\theta(\partial_{\phi}+in)
              - \frac{n\sin\theta}{1+\cos\theta} \right]   \\
J_{z} & = & -i(\partial_{\phi} + in) -n
\end{eqnarray}
which obeys the formal commutation relations
\begin{equation}
[{\bf J},{\cal H}] = 0 \; ; \;\; [J_{i},J_{j}]=i\varepsilon_{ijk}J_{k},
\end{equation}
for $\sin\theta \neq 0$.
We find
\begin{eqnarray}
J^{2} & = & n^{2} + \triangle   \nonumber \\
\Rightarrow \triangle & = & J^{2} - n^{2}.
\end{eqnarray}
Let us denote
\begin{equation}
J^{2}\Psi(\Omega) = l(l+1) \Psi(\Omega),
\end{equation}
where we set
\begin{equation}
\Psi(\Omega) = \exp[im\phi] P(\cos\theta) \; ; \;\; m=0,\pm 1,\pm 2, \cdots .
\end{equation}
Then
\begin{equation}
J_{z} \Psi(\Omega) = m \Psi(\Omega).
\end{equation}
The eigenvalues of $J^{2}$ are found to be given by $l(l+1)$ with
\begin{equation}
l \equiv k + \frac{1}{2} \left( |m+n|+|m-n| \right) \; ; \;\;
k=0,1,2,3,\cdots \; ; \;\; m=0,\pm 1,\pm 2, \cdots .
\end{equation}
The corresponding orthonormal eigenfunctions are
\begin{equation}
\Psi^{(k,m)}(\Omega) = C_{km}\left( \frac{1-\cos\theta}{2}\right)^{\frac{|m+n|}
{2}} \left( \frac{1+\cos\theta}{2} \right)^{\frac{|m-n|}{2}} \cdot
P^{(|m+n|,|m-n|)}_{k}(\cos\theta) e^{im\phi},
\end{equation}
where $P^{(a,b)}_{k}$ denote Jacobi polynomials, and 
\begin{equation}
C_{km} = \left( \frac{k!\Gamma(|m+n|+|m-n|+k+1)\Gamma(|m+n|+|m-n|+2k+1)}
                {4\pi\Gamma(|m+n|+k+1)\Gamma(|m-n|+k+1)} \right)^{\frac{1}{2}}.
\end{equation}
The complete spectrum, $\delta$, of the operator $\triangle$ is
\begin{eqnarray}
\delta & = & (k+\frac{1}{2}(|m+n|+|m-n|))(k+1+\frac{1}{2}(|m+n|+|m-n|)) 
              - n^{2}   \nonumber  \\
       &   & k=0,1,2,3, \cdots \; ; \;\; m=0,\pm 1,\pm 2, \cdots .
\end{eqnarray}
For $|m| \leq n$,
\begin{equation}
\delta = (n+k)^{2} - n^{2} + (n+k).
\end{equation}
For $|m| \geq n$,
\begin{equation}
\delta = (|m|+k)^{2} - n^{2} + (|m|+k).
\end{equation}
Thus,
\begin{equation}
\delta = (\max(n,|m|)+k)(\max(n,|m|)+k+1) - n^{2}.
\end{equation}
The ground state energy is generated when $ k=0$, $n \geq |m|$.
\begin{equation}
\delta = n(n+1) - n^{2} =n
\end{equation}
The corresponding ground state orthonormal eigenfunction has the form
\begin{equation}
\Psi^{(0,m)}(\Omega) = C_{0m} \left( \sin\frac{\theta}{2} \right)^{n+m}
                              \left( \cos\frac{\theta}{2} \right)^{n-m}
                       e^{im\phi}
\end{equation}
which is the same as that in Eq.~(\ref{psi-eigen}).

\begin{table}
\caption{The $n$th rank symmetric traceless tensors, $Q^{(n)}$, for $n=1,2,
3,4,$ and $6$ in 2-dimensional space.}
\label{n-tensor}
\begin{tabular}{|rcl|}  \hline
 $Q^{(1)}_{\alpha}$ & $=$ & $m_{\alpha}$  \\
 $Q^{(2)}_{\alpha_{1}\alpha_{2}}$ & $=$ & $m_{\alpha_{1}}m_{\alpha_{2}} 
      - \frac{1}{2}\delta_{\alpha_{1}\alpha_{2}}$   \\
 $Q^{(3)}_{\alpha_{1}\alpha_{2}\alpha_{3}}$ & $=$ & 
      $m_{\alpha_{1}}m_{\alpha_{2}}m_{\alpha_{3}} - 
      \frac{1}{4}(\delta_{\alpha_{1}\alpha_{2}}m_{\alpha_{3}} +
      \delta_{\alpha_{2}\alpha_{3}}m_{\alpha_{1}}+
      \delta_{\alpha_{3}\alpha_{1}}m_{\alpha_{2}})$  \\
 $Q^{(4)}_{\alpha_{1}\alpha_{2}\alpha_{3}\alpha_{4}}$ & $=$ & 
      $m_{\alpha_{1}}m_{\alpha_{2}}m_{\alpha_{3}}m_{\alpha_{4}}
      - \frac{1}{6}(\delta_{\alpha_{1}\alpha_{2}}m_{\alpha_{3}}m_{\alpha_{4}}
      + 5 \mbox{ more terms by permutations})$    \\
   & & $+\frac{1}{24}(\delta_{\alpha_{1}\alpha_{2}}
      \delta_{\alpha_{3}\alpha_{4}} 
      + 2 \mbox{ more terms by permutations})$    \\
 $Q^{(6)}_{\alpha_{1}\alpha_{2}\alpha_{3}\alpha_{4}\alpha_{5}\alpha_{6}}$ 
   & $=$ & $m_{\alpha_{1}}m_{\alpha_{2}}m_{\alpha_{3}}m_{\alpha_{4}}
   m_{\alpha_{5}}m_{\alpha_{6}} - \frac{1}{10}(\delta_{\alpha_{1}\alpha_{2}}
   m_{\alpha_{3}}m_{\alpha_{4}}m_{\alpha_{5}}m_{\alpha_{6}}
   + 14 \mbox{ more terms by permutations})$   \\
& & $+\frac{1}{80}(\delta_{\alpha_{1}\alpha_{2}}\delta_{\alpha_{3}\alpha_{4}}
   m_{\alpha_{5}}m_{\alpha_{6}}
   + 44 \mbox{ more terms by permutations})$  \\   
& & $-\frac{1}{480}(\delta_{\alpha_{1}\alpha_{2}}\delta_{\alpha_{3}\alpha_{4}}
   \delta_{\alpha_{5}\alpha_{6}}
   + 14 \mbox{ more terms by permutations})$  \\   \hline
 $m_{\alpha}$      & $=$ & $Q^{(1)}_{\alpha}$  \\
 $m_{\alpha_{1}}m_{\alpha_{2}}$ & $=$ & $Q^{(2)}_{\alpha_{1}\alpha_{2}} + 
                             \frac{1}{2}\delta_{\alpha_{1}\alpha_{2}}$   \\
 $m_{\alpha_{1}}m_{\alpha_{2}}m_{\alpha_{3}}$ & $=$ & 
    $Q^{(3)}_{\alpha_{1}\alpha_{2}\alpha_{3}} + 
    \frac{1}{4}(\delta_{\alpha_{1}\alpha_{2}}Q^{(1)}_{\alpha_{3}}+
    \delta_{\alpha_{2}\alpha_{3}}Q^{(1)}_{\alpha_{1}}+
    \delta_{\alpha_{3}\alpha_{1}}Q^{(1)}_{\alpha_{2}})$  \\
 $m_{\alpha_{1}}m_{\alpha_{2}}m_{\alpha_{3}}m_{\alpha_{4}}$ & $=$ & 
    $Q^{(4)}_{\alpha_{1}\alpha_{2}\alpha_{3}\alpha_{4}}
    + \frac{1}{6}(\delta_{\alpha_{1}\alpha_{2}}Q^{(2)}_{\alpha_{3}\alpha_{4}}
    + 5 \mbox{ more terms by permutations})$   \\
  & & $+\frac{1}{8}(\delta_{\alpha_{1}\alpha_{2}}\delta_{\alpha_{3}\alpha_{4}}
    + 2 \mbox{ more terms by permutations})$   \\
 $m_{\alpha_{1}}m_{\alpha_{2}}m_{\alpha_{3}}m_{\alpha_{4}}m_{\alpha_{5}}
 m_{\alpha_{6}}$ & $=$ & $Q^{(6)}_{\alpha_{1}\alpha_{2}\alpha_{3}\alpha_{4}
 \alpha_{5}\alpha_{6}} + \frac{1}{10}(\delta_{\alpha_{1}\alpha_{2}}
 Q^{(4)}_{\alpha_{3}\alpha_{4}\alpha_{5}\alpha_{6}}
 + 14 \mbox{ more terms by permutations})$  \\
  & & $ + \frac{1}{48}(\delta_{\alpha_{1}\alpha_{2}}
      \delta_{\alpha_{3}\alpha_{4}}Q^{(2)}_{\alpha_{5}\alpha_{6}}
      + 44 \mbox{ more terms by permutations})$  \\ 
  & & $ + \frac{1}{48}(\delta_{\alpha_{1}\alpha_{2}}
      \delta_{\alpha_{3}\alpha_{4}}\delta_{\alpha_{5}\alpha_{6}} 
      + 14 \mbox{ more terms by permutations})$  \\  \hline
\end{tabular}
\end{table}

\input{psfig}
\newpage

\end{document}